\documentclass[twoside,onecolumn,11pt]{article}
\usepackage{geometry}
\usepackage{amsmath}

\usepackage[utf8]{inputenc}
\usepackage{lmodern}
\usepackage[english]{babel}
\usepackage{graphicx}
\usepackage{subfigure}
\usepackage{breakcites}
\usepackage{float}
\usepackage[authoryear]{natbib}
\usepackage{cite}
\usepackage{color}                      

\usepackage{caption}
\usepackage{multirow}
\usepackage[breaklinks=true]{hyperref}
\usepackage{url}
  
\usepackage{lscape}
\usepackage{longtable}
\DeclareUnicodeCharacter{2212}{-}
\usepackage{booktabs}

\usepackage{enumitem}

\chardef\us=`\_

\usepackage{authblk}
\title{\Large\bf{On the Radial and Longitudinal Variation of a Magnetic Cloud: ACE, Wind, ARTEMIS and Juno Observations}}
\author[1]{E.E. Davies\thanks{Corresponding author email: emma.davies12@imperial.ac.uk}}
\author[1]{R.J. Forsyth}
\author[2]{S.W. Good}
\author[2]{E.K.J. Kilpua}
\affil[1]{The Blackett Laboratory, Imperial College London, London, UK}
\affil[2]{Department of Physics, University of Helsinki, Helsinki, Finland}
\date{\small{Accepted for publication in \textit{Solar Physics}: 12 September 2020}}                    
\usepackage{fancyhdr}
\begin{document}

\maketitle

\vspace{-0.5cm}
\begin{abstract}
    
\normalfont\noindent{{We present observations of the same magnetic cloud made near Earth by the Advance Composition Explorer (ACE), Wind, and the Acceleration, Reconnection, Turbulence and Electrodynamics of the Moon's Interaction with the Sun (ARTEMIS) mission comprising the Time History of Events and Macroscale Interactions during Substorms (THEMIS) B and THEMIS C spacecraft, and later by Juno at a distance of 1.2 AU. The spacecraft were close to radial alignment throughout the event, with a longitudinal separation of $3.6^{\circ}$ between Juno and the spacecraft near Earth. The magnetic cloud likely originated from a filament eruption on 22 October 2011 at 00:05 UT, and caused a strong geomagnetic storm at Earth commencing on 24 October. Observations of the magnetic cloud at each spacecraft have been analysed using Minimum Variance Analysis and two flux rope fitting models, Lundquist and Gold-Hoyle, to give the orientation of the flux rope axis. We explore the effect different trailing edge boundaries have on the results of each analysis method, and find a clear difference between the orientations of the flux rope axis at the near-Earth spacecraft and Juno, independent of the analysis method. The axial magnetic field strength and the radial width of the flux rope are calculated using both observations and fitting parameters and their relationship with heliocentric distance is investigated. Differences in results between the near-Earth spacecraft and Juno are attributed not only to the radial separation, but to the small longitudinal separation which resulted in a surprisingly large difference in the in situ observations between the spacecraft. This case study demonstrates the utility of Juno cruise data as a new opportunity to study magnetic clouds beyond 1 AU, and the need for caution in future radial alignment studies.}}

\vspace{0.5cm}

\noindent{\textbf{Keywords:} Coronal Mass Ejections, Interplanetary; Magnetic Clouds; Multi-spacecraft Observations; Radial Evolution; Longitudinal Variation; Juno.}
    
\end{abstract}

\section{Introduction}\label{sec:intro}

Interplanetary coronal mass ejections \citep[ICMEs: e.g.][]{kilpua2017coronal} are large scale structures of plasma and magnetic field that are driven from the solar atmosphere and propagate through the heliosphere. These transient structures are distinguished from the ambient solar wind in situ by features that may include an enhanced magnetic field, low plasma $\beta$, declining velocity profile and decreased proton and electron temperature amongst many other possible features \citep[e.g.][]{zurbuchen2006situ}. ICMEs with a strong and sustained southward magnetic field component are known to be the main drivers of strong geomagnetic activity at Earth \citep[]{gonzalez1987criteria, tsurutani1997interplanetary, echer2004geoeffectiveness, zhang2007solar, eastwood2008science, kilpua2017b} and therefore their evolution is of great interest in space weather forecasting.

Magnetic clouds are a subset of ICMEs which feature signatures including an enhanced magnetic field, smooth rotation of the magnetic field vector, low plasma $\beta$ and a drop in proton temperature \citep{burlaga1981magnetic}. Magnetic clouds exhibit well structured magnetic fields consistent with force-free flux ropes \citep{goldstein1983field} which comprise nested helical magnetic field lines wound around a central axis. The proportion of ICMEs that can be identified as magnetic clouds is on average $\approx$ 30\% \citep{gosling1990coronal}, but this varies with the solar cycle: at solar minimum, 60\% of ICMEs can be identified as magnetic clouds, whereas at solar maximum, this falls to 15\% \citep{cane2003interplanetary}.

To understand the evolution of ICMEs as they move out in the solar wind, it is useful to track signatures of specific events over different heliocentric distances. There are unfortunately a very limited amount of cases where multiple spacecraft that are at different heliospheric distances, but near radially aligned, have encountered the same ICME. Studies that describe such encounters to analyse ICME evolution include \citet{burlaga1981magnetic}, \citet{cane1997helios}, \citet{bothmer1997structure}, \citet{liu2008reconstruction}, \citet{mostl2009linking}, \citet{mostl2009optimized}, \citet{rouillard2010white}, \citet{farrugia2011}, \citet{nakwacki2011dynamical}, \citet{kilpua2011multipoint}, \citet{nieves2012remote}, \citet{ruffenach2012multispacecraft}, \citet{nieves2013inner}, \citet{good2015radial}). There has been a number of case studies \citep{winslow2016longitudinal, good2018correlation, kilpua2019, lugaz2019evolution} and statistical studies \citep{good2019self, vrvsnak2019heliospheric, salman2020radial} that have greatly expanded the number of analysed events in recent times.

The studies mentioned above have primarily used spacecraft at or within 1 AU. Radial alignment studies of ICMEs beyond Earth are particularly rare. One such study by \citet{mulligan1999intercomparison} compared four ICME events observed by the Near Earth Asteroid Rendezvous (NEAR) and Wind spacecraft, with radial separations between 0.18 and 0.63 AU and longitudinal separations between 1.2 and 33.4$^{\circ}$. However, with the introduction of more spacecraft into the solar wind in recent years, including planetary mission spacecraft during their cruise phase and/or outside of their respective planetary environments, there have been more opportunities for radial alignments between spacecraft and at larger heliospheric distances. The NASA Juno mission was launched in August 2011 with the science goals of exploring the origin and evolution of Jupiter \citep{bolton2017juno}. Juno cruise data, namely the magnetic field measured by the fluxgate magnetometer (MAG) between 2011 and 2016, provides a new opportunity to study ICME evolution beyond 1 AU, and is a key resource in understanding the chain of evolution of ICMEs through the heliosphere.

We present observations and analysis of an ICME with a clear magnetic cloud structure registered during 24 - 26 October 2011 by the Advance Composition Explorer (ACE), Wind, and the Acceleration, Reconnection, Turbulence and Electrodynamics of the Moon's Interaction with the Sun (ARTEMIS) mission in the near-Earth environment, and Juno at a heliocentric distance of 1.24 AU shortly after commencing its cruise phase to Jupiter. The near-Earth spacecraft and Juno were separated longitudinally by a maximum angle of just 3.6$^{\circ}$, with a maximum separation in latitude of only 0.1$^{\circ}$. This ICME is of particular interest as it caused the strongest geomagnetic storm at Earth in 2011, peaking at a Dst of -147 nT, driven by the southward magnetic fields preceding the magnetic cloud rather than the magnetic cloud itself where the magnetic fields were northward.

In this study we use observations from the multiple near-Earth spacecraft to provide several independent fits indicating the degree of variability of the fits along different trajectories through the ICME, and also to determine the direction of propagation of structures such as the ICME shock using the timing at each spacecraft, as discussed in Section \ref{sec:analysis}. The performance of the force-free fitting models is compared and explored for the different trailing edge boundaries chosen. Previous studies investigating model performance and the importance of flux rope boundary selection include \citet{riley2004fitting}, \citet{dasso2006new}, \citet{al2013magnetic,al2018fitting}, \citet{janvier2015comparing}. The flux rope orientations and other kinematic properties of the ICME resulting from the fitting techniques in this study are compared between the near-Earth spacecraft and Juno to analyse the evolution of the ICME.

\section{Spacecraft Observations} \label{sec:scobservations}

In situ observations made by the ACE, Wind, ARTEMIS, and Juno spacecraft are presented. Both ACE and Wind are NASA spacecraft that orbit the L1 Lagrangian point, upstream of the Earth. Wind was launched in November 1994, three years prior to ACE in August 1997. The ARTEMIS spacecraft comprise the Time History of Events and Macroscale Interactions during Substorms (THEMIS) B and THEMIS C spacecraft, two of the five THEMIS spacecraft launched in February 2007, and moved to a lunar orbit in 2010. The Juno mission was launched in August 2011 with the purpose of studying the magnetosphere and atmosphere of Jupiter, and reached Jupiter in July 2016. The five year cruise phase to 5 AU presents a new opportunity to study ICMEs beyond 1 AU. 

\begin{figure}
\centering
\includegraphics[width = 12cm]{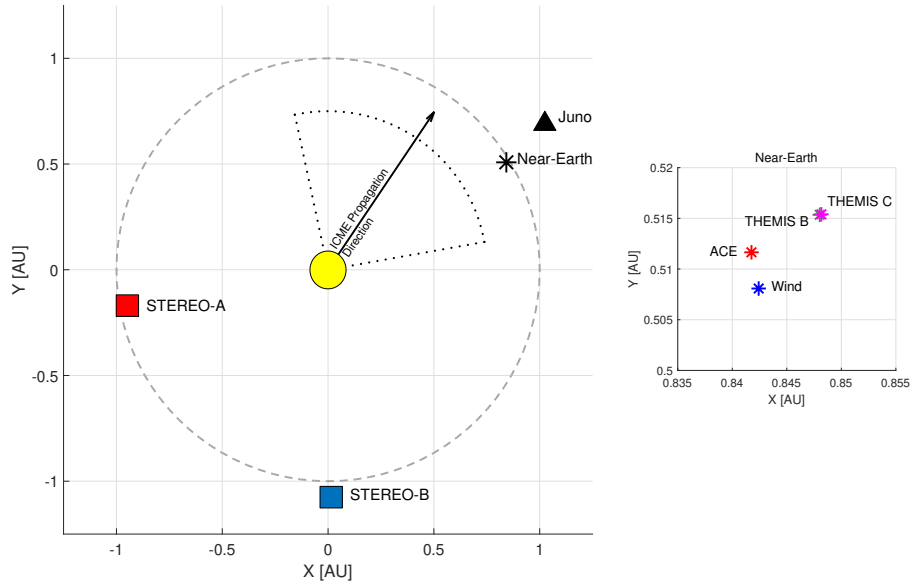}
\captionsetup{font=small, labelfont=bf}
\caption{Location of the near-Earth spacecraft (ACE, Wind, and ARTEMIS), Juno, STEREO-A and STEREO-B on 25 October 2011 at 00:00 UT in Heliocentric Aries Ecliptic (HAE) coordinates. Imagers onboard the STEREO spacecraft can be used to estimate the initial propagation direction and extent of the ICME, represented by the black arrow and dotted lines, respectively (for details, see Section \ref{sec:context}). The near-Earth spacecraft are in near radial alignment with Juno, with a small maximum longitudinal separation of 3.6$^{\circ}$. The inset shows a close up of the configuration of the near-Earth spacecraft: ACE (red), Wind (blue), THEMIS B (green) and THEMIS C (magenta). The near-Earth spacecraft configuration has a maximum radial separation of $<0.01$ AU and a maximum longitudinal separation of $<0.2^{\circ}$.}
\label{fig:orbitplot}
\end{figure}

The spacecraft positions of ACE, Wind, and ARTEMIS (denoted as `Near-Earth'), the Solar Terrestrial Relations Observatory (STEREO)-A, STEREO-B, and Juno on 25 October 2011 at 00:00 UT are shown in Figure \ref{fig:orbitplot}, using Heliocentric Aries Ecliptic (HAE) coordinates. Figure \ref{fig:orbitplot} demonstrates that Juno and the near-Earth spacecraft were in near radial alignment, with a maximum longitudinal separation of just 3.6$^{\circ}$. The near-Earth spacecraft configuration itself has a maximum radial separation of $<0.01$ AU and a maximum longitudinal separation of $<0.2^{\circ}$.

To identify the ICME in situ near Earth, we use measurements of the magnetic field taken by the magnetometers onboard ACE \citep[Magnetic Field Experiment, MAG:][]{smith1998ace}, Wind \citep[Magnetic Field Investigation, MFI:][]{lepping1995wind}, and the ARTEMIS mission \citep[Fluxgate Magnetometer, FGM:][]{auster2008themis}. Measurements of the solar wind plasma were used at ACE \citep[Solar Wind Electron Proton Alpha Monitor, SWEPAM:][]{mccomas1998solar}, Wind \citep[Solar Wind Experiment, SWE:][]{ogilvie1995swe} and ARTEMIS \citep[Electrostatic Analyzer, ESA:][]{mcfadden2008themis} to aid in the identification of the ICME in the near-Earth environment. To identify the ICME at Juno, only measurements of the magnetic field were used \citep[Magnetic Field Experiment, MAG:][]{connerney2017juno}, as the plasma experiment onboard Juno was not turned on in October 2011 (it was first turned on during the final month of approach to Jupiter for calibration).

\begin{figure}
\centering
\includegraphics[width=\textwidth]{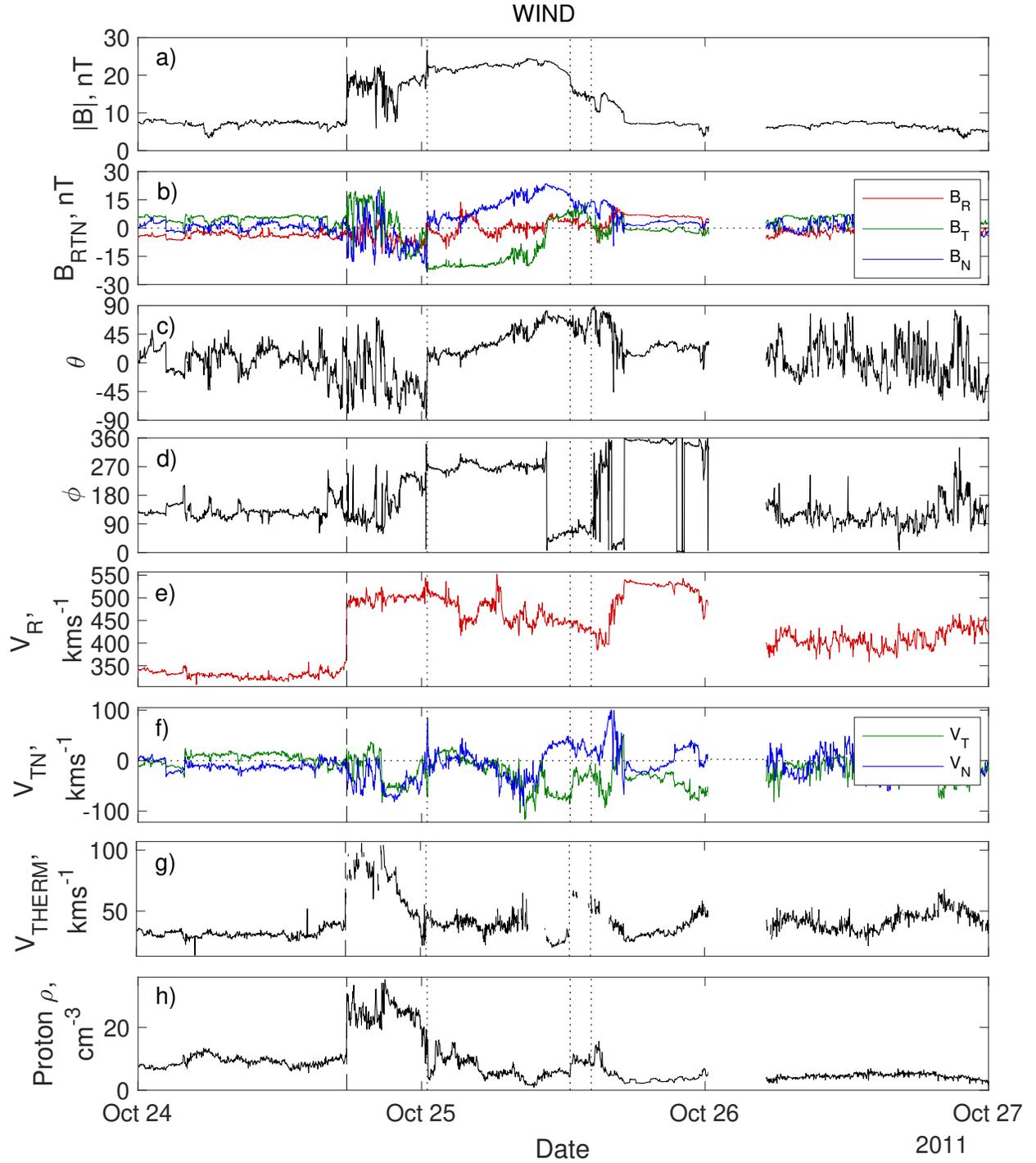}
\captionsetup{font=small, labelfont=bf}
\caption{In situ magnetic field (1 minute resolution) and plasma (1 minute 38 second resolution) signatures observed by Wind. The vertical dashed line indicates the shock and the vertical dotted lines indicate the boundaries of the flux rope, with two possible trailing edge locations. The respective panels display a) the total magnetic field, b) the components of the magnetic field in RTN coordinates (the radial component is shown in red, the transverse component in green, and the normal component in blue), c) the calculated $\theta$, and d) $\phi$ angles of the magnetic field, e) the radial proton speed, f) the transverse and normal components of the proton velocity, g) the thermal proton velocity, and h) proton density.}
\label{fig:wind}
\end{figure}

Figure \ref{fig:wind} presents the magnetic field (1 minute resolution) and plasma (1 minute 38 second resolution) signatures observed at Wind. These observations are very similar in the large-scale to those at ACE and both ARTEMIS spacecraft (in situ signatures observed by these spacecraft can be found in the electronic supplementary material: Figures 7, 8 and 9), and are therefore representative of the near-Earth environment. The structure delineated by the dotted vertical lines displays features associated with an ICME such as the enhancement of the magnetic field (panel a), declining radial speed profile (panel e), and the decrease in both proton temperature (panel g) and density (panel h), which distinguish it from the ambient solar wind. A shock (vertical dashed line) driven by the ICME was observed at Wind at 17:40 UT on 24 October 2011. Table 1 lists the shock arrival time, $t_S$, at each of the near-Earth spacecraft and the heliocentric distance, $r_H$, at which the spacecraft were located, which are used in Section \ref{subsec:timings} to infer the propagation direction of the shock. Figure \ref{fig:wind} also displays the magnetic field components in radial tangential normal (RTN) coordinates (panel b), the angle of the magnetic field vector to the R-T plane, $\theta$ (panel c), and the angle of the magnetic field vector swept out anticlockwise from the Sun-Earth line, projected onto the R-T plane, $\phi$ (panel d). The ICME meets the criteria detailed by \citet{burlaga1981magnetic} to classify magnetic clouds: a strong enhancement of the magnetic field, smooth rotation of the magnetic field components, a low variance of the magnetic field, and low proton temperature and density. 

Identifying the leading and trailing edges of the magnetic cloud flux rope can often be subjective and features are not always coincident \citep[e.g. see discussion in][]{richardson2010,kilpua2013}. In the case of this event, we identify two possible trailing edges: the first coincides with the earliest significant drop in magnetic field strength and a slight increase in proton temperature and density, the second coincides with the end of both the smooth magnetic field rotation and the declining radial speed profile. The leading edge of the flux rope is easier to identify, marking the start of the magnetic field enhancement, smooth rotation of the magnetic field, and the steady decline in radial speed. There is also a short substructure featured by a dip in the magnetic field magnitude (approximately 10 minutes in duration) at the leading edge of the ejecta observed at each spacecraft. Such substructures are generally reported at the leading edges of magnetic clouds and are thus a solid indicator of the boundary \citep{wei2003}. These substructures can result from the interaction between the magnetic cloud and the preceding solar wind, or be relics of the CME release process at the Sun \citep{farrugia2001,kilpua2013}.

We define the sheath of the ICME as the region between the shock front and the leading edge of the flux rope. In Figure \ref{fig:wind}, the sheath region displays a variable and fluctuating magnetic field structure at Wind, followed by a region of low variance which begins at approximately 22:00 UT. This region of low variance is also observed across each of the near-Earth spacecraft and coincides with a change in electron pitch angle signature at ACE \sloppy{(not shown; available at \url{http://www.srl.caltech.edu/ACE/ASC/DATA/level3/swepam/swepam_pa_summary/2011-297-12.png}) where two oppositely directed strahls appear at both 0$^{\circ}$ and 180$^{\circ}$ pitch angles, meaning that counterstreaming electrons are present, a feature often indicative of closed magnetic field lines associated with an ICME \citep{gosling1990coronal}. The onset of the counterstreaming electron flows, which extend through the flux rope, indicates the true start of the ICME (but not the flux rope itself), and is coincident with the fall of the proton temperature.}

The radial component of the proton velocity in Figure \ref{fig:wind} displays a declining speed profile from 527 kms$^{-1}$ to 455 or 425 kms$^{-1}$, trailing edge boundary dependent, during this period indicating the expansion of the magnetic cloud. Following the flux rope, there is a steep increase in the radial component of the velocity to 540 kms$^{-1}$, and clear velocity deflections in the transverse and normal directions of -93 and 100 kms$^{-1}$, respectively. This feature is observed by each of the near-Earth spacecraft and is also reflected in the magnetic field data by a compressed region of magnetic field following the later trailing edge boundary, preceding features consistent with a reverse shock. It is possible that the compressed magnetic field region was the result of a weak ICME following the event as the magnetic field has a low variance and temperature, or a consequence of a transient coronal hole opened by the CME related to the magnetic cloud \citep{luhmann1998relationship}. The dip in magnetic field at the later trailing edge of the flux rope is in magnetic and thermal pressure balance (not shown), indicating that this is likely the true ICME boundary. Previous studies have also defined the trailing edge to be this later boundary \citep{lepping2015wind,wood2017stereo}, whilst \citet{nieves2019unraveling} define the ICME as a complex structure and extend the boundary to the end of the region of increased radial velocity. It is probable that the cause of the ambiguity in trailing edge selection is due to the solar wind following the magnetic cloud having led to a compression and heating of the trailing part of the flux rope that moved the temperature increase forward into the flux rope, affecting observed features until the earlier trailing edge defined.

\begin{figure*}
\centering
\includegraphics[width=\textwidth]{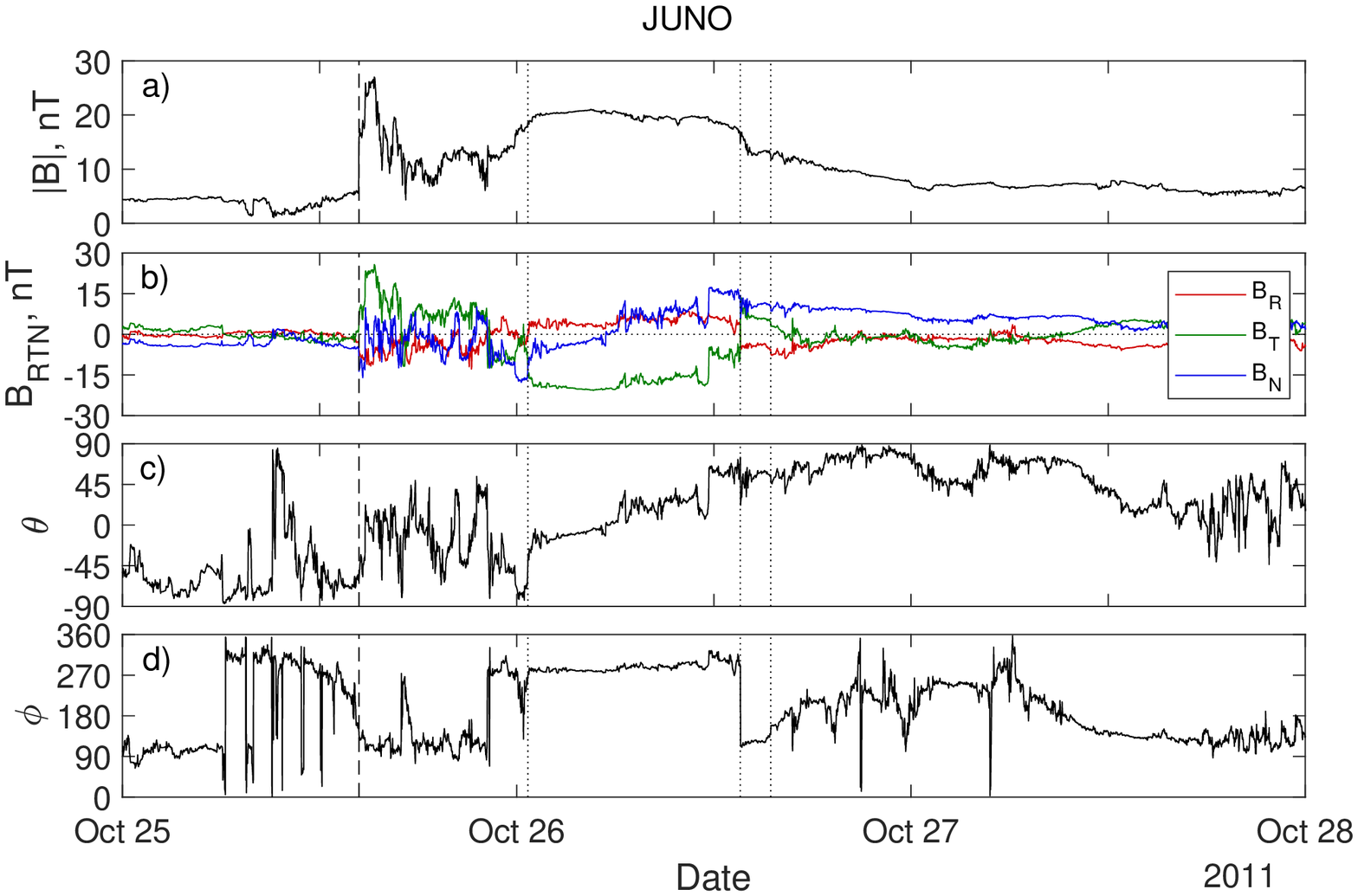}
\captionsetup{font=small, labelfont=bf}
\caption{In situ magnetic field (1 minute resolution) signatures observed by Juno displayed in the same format as Figure \ref{fig:wind}.}
\label{fig:juno}
\end{figure*}

Figure \ref{fig:juno} presents the magnetic field (1 minute resolution) signatures observed by Juno, in the same format as Figure \ref{fig:wind}. The structure observed is very similar to that of Figure \ref{fig:wind}, with an enhanced magnetic field (delineated by the vertical dotted lines) and similar behaviour of the magnetic field components during this enhancement, which display a clear flux rope structure. A shock front is registered by Juno at a heliocentric distance of 1.24 AU on 25 October at 14:23 UT, less than a day later than the shock at Wind. Assuming a constant velocity and radial propagation, the timing of the shock front between Wind and Juno gives an average velocity of 518 kms$^{-1}$, in reasonable agreement with the observed shock velocity of 489 kms$^{-1}$ at Wind. Much of the sheath magnetic field is lower in magnitude in the Juno observations except for an initial increase behind the shock front in comparison to observations at Wind, and the low variance region is harder to distinguish. The duration of the sheath is longer at Juno than at Wind, suggesting that it has expanded as the ICME has propagated, although the difference could also be due to spatial variation at different measurement locations \citep{kilpua2017coronal}. The mean magnetic field enhancement within the flux rope is 19.7/18.9 nT (depending on trailing edge definition) at Juno in comparison to 22.4/21.5 nT at Wind. The format of `earlier/later trailing edge' is used to present dependent values throughout. The mean magnetic field values at each spacecraft are summarised in Table 1. The magnitude profiles, however, follow different trends: at Wind the magnitude slowly increases between the leading and earlier trailing edge, whereas the profile at Juno decreases slightly over the same boundaries. For faster ICMEs, the magnetic field profiles are often asymmetric, with a larger magnetic field strength at the front of the flux rope than the back \citep{janvier2019generic}, as we observe at Juno. This can be considered as an effect of the time difference in observations between the leading and trailing edges of the flux rope where, in an expanding flux rope, the magnetic field weakens in the time taken to pass the observing spacecraft. Whilst the assumed correlation between this asymmetric magnetic field profile and expanding structures tends to hold true, \citet{nieves2018understanding} found in a study of 298 ICMEs with well structured magnetic topologies that 22$\%$ of positively expanding structures had compression at the back of the flux rope, and suggested that this could be an  effect of the curvature of the passing structure. The observations at Wind show a declining radial speed profile, and therefore a positive expansion of the flux rope. We suggest in this case that the increasing magnetic field profile at Wind is likely a product of magnetic field compression due to the increased radial speed of the solar wind following the flux rope. However, this compression is not observed at Juno and therefore, a more typical magnetic field profile is observed. The difference in the magnetic field components within the flux rope boundaries between Wind and Juno can be seen by comparing Figures \ref{fig:wind} and \ref{fig:juno}: the normal component is similar in profile yet differs in value between the two heliocentric distances changing from negative (south) to positive (north) at Juno but remaining north at Wind, while the radial component shows more significant dissimilarities. The transverse component is the only component to remain of a similar shape and magnitude between Wind and Juno. The field angles are also interesting, as although very similar in profile, we note that the discontinuity in $\phi$ occurs later within the rope at Juno than at Wind relative to the trailing edges of the flux rope. By studying the magnetic field components and how they evolve throughout the flux rope, we can obtain a sense of handedness. Using the classification system following \citet{bothmer1997structure} and \citet{mulligan1998solar}, the flux rope can be classified as either SEN at Juno, or ENW at Wind. SEN means that at the leading edge of the flux rope the field points to the south, then rotates to point east at the axis and finally rotates to north at the trailing edge. Similarly for the ENW classification, the leading edge points to the east, rotates to point north at the axis, and finally rotates to point west at the trailing edge. Both classifications are left-handed. The consistency of the handedness is supporting evidence that the spacecraft observed the same event, as the handedness of a flux rope remains the same as it propagates \citep{marubashi2015geometrical}. Following the flux rope, the clear drop in field magnitude observed by Wind is not present in the Juno observations, although there is a region of modestly enhanced but declining magnitude. This decrease is smoother at Juno implying expansion after the trailing edge, although this may again be due to a difference in measurement location.

The arrival times of the shock front, $t_S$, and the flux rope leading, $t_L$, and trailing edges, $t_{T1}$ and $t_{T2}$, observed at each spacecraft are presented in rows 2\textendash5 of Table 1. The difference between the earlier/later trailing edge times and the leading edge time is consistent with the expansion of the flux rope as its duration is observed to be 12 hours and 14 minutes/13 hours and 53 minutes at Wind and 12 hours and 56 minutes/14 hours and 47 minutes when observed at Juno. However, the difference in duration may also occur due to the different spacecraft trajectories through the ICME and the potentially different ICME propagation speeds at each spacecraft. The observed radial velocities at each boundary, $v_L$ and $v_T$, are presented in rows 7 and 8 of Table 1, where there are two values for $v_T$ as the parameter is dependent on the trailing edge used. The leading and trailing edge velocities are consistent for each of the near-Earth spacecraft. Row 9 of Table 1 presents the expansion velocity, $v_E$, calculated as half of the difference between the trailing and leading edge velocities. The mean expansion velocity, $\langle v_E \rangle$, is presented in row 11 of Table 1. It takes into account the timings and heliocentric distance between the leading and trailing edges at each of the near-Earth spacecraft and Juno to give the mean propagation speed of the leading and trailing edges. $\langle v_E \rangle$ is calculated as half the difference between these speeds. The mean expansion velocity between Wind and Juno was found to be 4.8/5.1 kms$^{-1}$; much smaller than the observed expansion velocity at Wind of 36/51 kms$^{-1}$. This indicates a slowing of the expansion velocity as the ICME propagates. 

\begin{table}[t]
\resizebox{\textwidth}{!}{%
\begin{tabular}{@{}cccccc@{}}
\toprule
\multicolumn{1}{l}{} & Wind & ACE & THEMIS B & THEMIS C & Juno \\ \midrule
\multicolumn{1}{c|}{$r_H$ {[}AU{]}} & 0.984 & 0.985 & 0.992 & 0.993 & 1.24 \\
\multicolumn{1}{c|}{$t_S$ (2011)} & Oct 24 17:40 UT & Oct 24 17:48 UT & Oct 24 18:43 UT & Oct 24 18:44 UT & Oct 25 14:23 UT \\
\multicolumn{1}{c|}{$t_L$ (2011)} & Oct 25 00:28 UT & Oct 25 00:35 UT & Oct 25 01:12 UT & Oct 25 01:12 UT & Oct 26 00:40 UT \\
\multicolumn{1}{c|}{$t_{T1}$ (2011)} & Oct 25 12:34 UT & Oct 25 12:42 UT & Oct 25 13:23 UT & Oct 25 13:23 UT & Oct 26 13:36 UT \\
\multicolumn{1}{c|}{$t_{T2}$ (2011)} & Oct 25 14:21 UT & Oct 25 14:21 UT & Oct 25 15:04 UT & Oct 25 15:04 UT & Oct 26 15:27 UT \\
\multicolumn{1}{c|}{$v_S$ {[}kms$^{-1}${]}} & 487 & 475 & 486 & 484 & - \\
\multicolumn{1}{c|}{$v_L$ {[}kms$^{-1}${]}} & 527 & 494 & 485 & 496 & - \\
\multicolumn{1}{c|}{$v_{T}$ {[}kms$^{-1}${]}} & 455/425 & 442/426 & 440/414 & 451/415 & - \\
\multicolumn{1}{c|}{$ v_{E}$ {[}kms$^{-1}${]}} & 36/51 & 26/34 & 22/36 & 43/41 & - \\
\multicolumn{1}{c|}{$ v_c$ {[}kms$^{-1}${]}} & 472/468 & 462/458 & 449/442 & 443/440 & - \\
\multicolumn{1}{c|}{$ \langle v_{E} \rangle$ {[}kms$^{-1}${]}} & 4.8/5.1 & 4.7/6.1 & 4.1/5.2 & 4.1/5.1 & - \\
\multicolumn{1}{c|}{$\langle v_c \rangle $ {[}kms$^{-1}${]}} & 490/476 & 468/460 & 463/449 & 473/455 & - \\
$D$ {[}AU{]} & 0.137/0.156 & 0.135/0.152 & 0.132/0.148 & 0.130/0.147 & 0.147/0.164 \\
\multicolumn{1}{c|}{$B_{max}$ {[}nT{]}} & 26.7 & 25.0 & 26.9 & 25.4 & 21.0 \\
\multicolumn{1}{c|}{$\langle B \rangle$ {[}nT{]}} & 22.4/21.5 & 22.7/21.8 & 23.0/22.1 & 22.8/22.0 & 19.7/18.9 \\ \bottomrule
\end{tabular}%
}
\captionsetup{font=small, labelfont=bf}
\caption{Shock and flux rope parameters of the ICME observed at each spacecraft, including the heliocentric distance of the spacecraft ($r_H$) and the times at which the shock front, leading and trailing edges were observed ($t_S$, $t_L$, $t_{T1}$, $t_{T2}$, respectively). Shock front ($v_S$), leading edge ($v_L$), trailing ($v_{T}$), expansion ($ v_{E}$), and cruise velocities ($ v_c$) have been calculated for the near-Earth spacecraft where plasma data is available. The mean expansion ($ \langle v_{E} \rangle$) and cruise ($\langle v_c \rangle $) velocities have been calculated considering the propagation times between the near-Earth spacecraft and Juno. The observed radial width of the flux rope ($D$), and the maximum and mean magnetic field magnitudes observed inside the flux rope are given ($\langle B \rangle$ and $B_{max}$, respectively). Two values separated by `/' are presented where changing between trailing edge times, $t_{T1}$ and $t_{T2}$, affects the parameters.}
\label{tab:timings}
\end{table}

The observed radial width, $D$, can be calculated considering the cruise velocity of the flux rope and the time taken for a spacecraft to traverse the flux rope: $D = v_c (t_T-t_L)$. Here, the cruise velocity for each individual spacecraft is taken to be the solar wind velocity at the mid-point of the flux rope \citep{owens2005characteristic} and is used as an approximation of the average propagation speed of the magnetic cloud. The cruise velocity, $v_c$, is noted in Table 1 for each of the near-Earth spacecraft. As there is no plasma data for Juno during this period, the mean cruise speed, $\langle v_c \rangle$, has been calculated between each of the near-Earth spacecraft and Juno using timing considerations of both leading and trailing edges observed. The mean of these values, 473.5/460.0 kms$^{-1}$, has been taken as the cruise velocity used to calculate the radial width of the flux rope at Juno. The calculated radial widths are also given in row 13 of Table 1. The radial width of the flux rope was calculated to be 0.137/0.156 AU at Wind and 0.147/0.164 AU at Juno, with associated errors of approximately $\pm 0.004$ AU. The calculated widths are less than the average width of a flux rope at 1 AU of approximately 0.2 AU \citep{bothmer1997structure, liu2005statistical, gulisano2010global}. Using the boundary times defined and expansion velocities given by Table 1, one would expect an expansion of 0.019/0.024 AU between Wind and Juno, trailing edge dependent. We actually observe an expansion of the flux rope of 0.010/0.008 AU between Wind and Juno which is less than expected. The calculation of observed radial width does not take into account the orientation of the flux rope and therefore, is in real terms the length of the spacecraft trajectory through the rope. The mean cruise velocity of the ICME at each near-Earth spacecraft was also used as an approximate cruise velocity at Juno, and therefore it is perhaps unsurprising that the expected expansion is not observed. 

The maximum magnetic field strength within the flux rope, $B_{max}$, and the mean magnetic field strength within the flux rope, $\langle B \rangle$, are given in rows 14 and 15 in Table 1, respectively. The mean magnetic field magnitudes decrease between the near-Earth spacecraft and Juno as $\langle B \rangle \propto r_H^{-0.63\pm0.04}$ for both trailing edge times defined, where $r_H$ is heliocentric distance. We also find that $B_{max} \propto r_H^{-0.94\pm0.23}$. Previous studies that derived the relationship between magnetic field strength and heliocentric distance beyond 1 AU include \citet{ebert2009bulk} who found $B \propto r_H^{−1.29 \pm 0.12}$ and \citet{richardson2014identification} who found $B \propto r_H^{−1.21 \pm 0.09}$. Both studies used magnetic field data from Ulysses between 1 and 5.4 AU and calculated the relationships using the mean magnetic field of the ICMEs studied. The $B_{max}$ relationship is most similar to these relationships, with a slight overlap in associated errors. The disagreement with the mean magnetic field relationship derived in this study is likely due to the differences in the magnetic fields along the different paths taken by the near-Earth spacecraft and Juno through the flux rope. These longitudinal effects dominate over the expected small change in the field intensity due to the radial separation of 0.24 AU.

\section{Solar Source of the In Situ Structures} \label{sec:context}

To give context to the in situ observations, we try to locate the solar counterpart of the investigated magnetic cloud. Using the leading edge speed of the ICME at Wind and assuming a constant propagation speed from the Sun to L1, we find an estimated eruption time on 21 October, at 19:15 UT. Two potential candidate CMEs are listed in the Coordinated Data Analysis Workshop (CDAW) Solar and Heliospheric Observatory (SOHO) Large Angle and Spectrometric Coronagraph (LASCO) CME catalogue (\url{https://cdaw.gsfc.nasa.gov/CME_list/}) around this time. These were first observed in the LASCO C2 telescope field of view at 01:25 UT and 10:24 UT on 22 October, with 2$^{nd}$-order speeds at 20 $R_s$ of 663 and 1074 kms$^{-1}$, respectively. Both CMEs were also seen by the coronagraphs onboard STEREO-A and STEREO-B. Using the STEREO CME Analysis Tool (StereoCAT; \url{https://ccmc.gsfc.nasa.gov/stereocat/}) we find that the apex of the first CME had an initial propagation direction of 25$^{\circ}$ longitude west of the Earth-Sun line and 50$^{\circ}$ latitude north of the solar ecliptic (SE) plane with a half-width of 46$^{\circ}$. The later CME had an initial propagation direction of 90$^{\circ}$ longitude and 52$^{\circ}$ latitude with a half-width of 55$^{\circ}$. Based on these initial propagation directions, it is therefore likely that the source of the transient observed in situ at Earth and Juno is the first CME listed, in agreement with the \citet{wood2017stereo} STEREO survey of ICMEs observed in situ at Earth. 

The identified CME is associated with a filament eruption, studied in detail by \citet{gosain2016interrupted}. The magnetic configuration of filaments is observed to be that of a flux rope (Guo et al. 2010). The filament was located in the solar northern hemisphere indicating that the flux rope should likely be left-handed \citep{rust1994spawning}, which is consistent with the ENW/SEN flux rope classifications observed in situ. The propagation direction of the filament eruption stabilised at approximately 15$^{\circ}$ longitude and 45$^{\circ}$ latitude \citep{gosain2016interrupted}, consistent with the ICME propagation direction found using the STEREO CME Analysis Tool.

To further confirm the solar CME counterpart of the in situ magnetic cloud, we compare the observed ICME arrival times with those predicted by the Propagation Tool developed at the Institute of Research in Astrophysics and Planetology (IRAP) (\citet{rouillard2017propagation}; \url{http://propagationtool.cdpp.eu/}). Inputting values to the Propagation Tool recorded by the CDAW SOHO LASCO CME catalogue with a background solar wind speed observed in situ at Wind of 320 kms$^{-1}$ and an approximate drag parameter of 0.2$\times10^{-7}$ km$^{-1}$ resulted in predicted arrival times of the flux rope leading edge of 25 October 2011 06:34 UT at Wind and 26 October 2011 04:58 UT at Juno. The drag parameter used was derived using solar wind and magnetic cloud densities in combination with the observed radial width of the flux rope at Wind as in \citet{cargill2004aerodynamic}. The predicted times compare well with the observed leading edge arrival times of 25 October 2011 00:28 UT and 26 October 00:40 UT at Wind and Juno, respectively.

The source of the sudden increase in radial speed following the flux rope observed in situ by each of the near-Earth spacecraft remains inconclusive. Inspection of Solar Dynamics Observatory data \citep[Atmospheric Imaging Assembly, AIA:][]{lemen2011atmospheric} shows a small coronal hole already present prior to the eruption, at a similar latitude close to the filament channel from which the source of the magnetic cloud originated. An ENLIL simulation at \url{https://iswa.gsfc.nasa.gov/downloads/20111022_072000_anim.tim-vel.gif} shows a slight narrow stream of higher speed solar wind around the time of the event at Earth, but care must be taken whether to trust such small features. The in situ data at Wind in Figure \ref{fig:wind} shows a region of low temperature following the flux rope which could be evidence of a weak ICME following the event. Although there are no suitable CMEs listed in the CDAW SOHO LASCO CME catalogue, many ICMEs observed at Earth do not have clearly identifiable associated solar counterparts \citep{richardson2010, kilpua2014solar}.

\section{Analysis} \label{sec:analysis}

We use five methods to analyse the ICME. These include using: (i) timing considerations between the near-Earth spacecraft to determine a direction of propagation of the ICME shock, minimum variance analysis to estimate (ii) the propagation direction of the sheath region of the ICME at each spacecraft and (iii) the orientation of the flux rope, and the (iv) Lundquist and (v) Gold-Hoyle force-free flux rope fitting methods to provide independent determinations of the flux rope axis orientation and other parameters such as the axial magnetic field strength, impact parameter, and radial width scale value. Table 2 summarises the results of each method, and Figure \ref{fig:orientations} presents a visualisation of the calculated directions/orientations in terms of $\theta$ and $\phi$ for each analysis method and trailing edge definition.

\subsection{Timing Considerations} \label{subsec:timings}

Assuming a constant propagation velocity, $\boldsymbol{V_s}$, and a planar shock front, we use timing considerations of the shock front between the four near-Earth spacecraft to calculate the normal direction, $\boldsymbol{n_s}$, and speed of the shock, $\nu$:

\begin{subequations}
  \begin{equation}
    \label{eq-R}
      (\boldsymbol{R_x} - \boldsymbol{R_1}) \cdot \boldsymbol{n_s} = \nu \cdot \Delta t_{x1},
  \end{equation}
  \begin{equation}
    \label{eq-V}
    \boldsymbol{V_s} = \nu \cdot \boldsymbol{n_s},
  \end{equation}
\end{subequations}

\noindent where $\boldsymbol{R_x} - \boldsymbol{R_1}$ is the position of a spacecraft (where $x = 2,3,4$) relative to the spacecraft at $\boldsymbol{R_1}$, and $\Delta t_{x1}$ is the difference in shock arrival times \citep{mostl2012multi}. The calculated shock propagation direction is listed in Table 2 as $\theta_p = -9.4^{\circ}$ with respect to the solar ecliptic (SE) plane and $\phi_p = 11.0^{\circ}$ with respect to the Sun-Earth line, and has a propagation velocity of 514 kms$^{-1}$. This direction is visually presented in the first row of Figure \ref{fig:orientations} (shown in red). The observed shock velocity at each near-Earth spacecraft is given in Table 1, the mean of which was calculated to be 483 kms$^{-1}$. The calculated propagation velocity is therefore in reasonable agreement with observations.

\subsection{Minimum Variance Analysis} \label{subsec:mva}

Minimum variance analysis (MVA) has been performed on both the sheath region and flux rope. The technique involves calculating the eigenvectors and eigenvalues of a covariance matrix of the magnetic field components. When applied to a planar magnetic structure \citep[PMS; as in][]{nakagawa1989planar, neugebauer1993origins} of the sheath region, the minimum variance eigenvector corresponds to the normal direction of the PMS \citep{paschmann1998analysis}. The normal to the PMS has been found to be in good agreement with the shock normal in sheath regions where the PMS is found close to the shock front \citep{palmerio2016planar}. When MVA is applied to a flux rope, the intermediate eigenvector corresponds to the direction of the flux rope axis \citep{goldstein1983field}. 

Table 2 summarises the results of the MVA, performed on the PMS within the sheath region that immediately follows the shock front and the flux rope, where $\theta_a$ is the elevation angle out of the SE plane and $\phi_a$ is the angle from the Sun-Earth line anticlockwise in the SE plane. Figure \ref{fig:orientations} presents the orientations calculated by MVA in black. The mean normal direction to the sheath region at the near-Earth spacecraft was calculated to be $\theta = -11.9^{\circ}$ and $\phi = 7.4^{\circ}$. Comparing the calculated sheath normal with the direction of propagation of $\theta = -9.4^{\circ}$ and $\phi = 11.0^{\circ}$ calculated in Section \ref{subsec:timings}, we find these to be consistent between the near-Earth spacecraft, shown in the first row of Figure \ref{fig:orientations}. The sheath normal at Juno was calculated to be $\theta = -22.4^{\circ}$ and $\phi = 21.0^{\circ}$ and therefore there is a mean change in direction of $\theta = 10.6^{\circ}$ away from the SE plane and $\phi = 13.6^{\circ}$ anticlockwise in the SE plane between the near-Earth spacecraft and Juno.

The flux rope orientations obtained are well defined considering that the ratios of the maximum eigenvalue, $\lambda_1$, and minimum eigenvalue, $\lambda_3$ to the intermediate eigenvalue, $\lambda_2$ (both also summarised in Table 2) meet the criteria defined by \citet{siscoe1972significance} of $\frac{\lambda_1}{\lambda_2} > 1.37$ and $\frac{\lambda_3}{\lambda_2} < 0.72$. The resulting flux rope orientations, given in Table 2 and presented in the second and third rows of Figure \ref{fig:orientations}, are consistent within errors at the near-Earth spacecraft, with a mean orientation of $\theta = 52.4/70.0^{\circ}$ and $\phi = 220.2/245.8^{\circ}$ and a standard deviation from the mean of $\theta = 7.0/3.9^{\circ}$ and $\phi = 11.1/4.6^{\circ}$, where the first result is calculated using the earlier trailing edge time and the second uses the later trailing edge time. The flux rope orientation at Juno was calculated to be $\theta = 23.7/49.9^{\circ}$ and $\phi = 290.1/317.3^{\circ}$. Between the near-Earth spacecraft and Juno, the flux rope orientations display a clear mean change of $\theta = 28.7/20.1^{\circ}$ towards the SE plane and $\phi = 69.9/71.5^{\circ}$ anticlockwise in the SE plane. Although the calculated flux rope orientations at the near-Earth spacecraft are more moderately-inclined than highly-inclined, they support the flux rope classification of ENW at the near-Earth spacecraft and a lower-inclination flux rope of SEN at Juno.

\subsection{Force-Free Flux Rope Fitting} \label{subsec:fffrf}

MVA was also used as a starting point for the first force-free flux rope model fit to the magnetic field components, based on the Lundquist solutions \citep{lundquist1950}:

\begin{subequations}
  \begin{equation}
    \label{eq-a}
      B_r = 0,
  \end{equation}
  \begin{equation}
    \label{eq-b}
    B_{\phi} = B_0 J_1(\alpha r),
  \end{equation}
  \begin{equation}
    \label{eq-c}
      B_z = B_0 J_0(\alpha r).
  \end{equation}
\end{subequations}
\\
These solutions assume a force-free magnetic field with a constant $\alpha$ in a cylindrical configuration, where $J_0$ and $J_1$ are the zeroth and first order Bessel functions, $B_0$ is the magnetic field strength along the axis, and $r$ is the radial distance from the rope axis. The magnetic field solutions were fitted to the data of ACE, Wind, THEMIS B, THEMIS C, and Juno using a least squares procedure similar to that developed by \citet{lepping1990magnetic} where the calculated MVA orientation initialises the $\chi^2$ minimisation; details of this technique are given in \citet{good2019self} and \citet{kilpua2019}.  

The other flux rope fitting method used considers the magnetic field to have a uniform twist across the rope cross-section: a `Gold-Hoyle' tube \citep{gold1960origin}. There has been recent interest \citep[e.g.][]{hu2014structures, hu2015magnetic, wang2016twists} in using the Gold-Hoyle model to fit flux ropes in situ. This has been motivated by evidence \citep{kahler2011solar} indicating that field line lengths, as estimated from strahl electron travel times from the Sun, are too short to be consistent with the highly twisted (and hence very long) field lines in the outer layers of a Lundquist flux rope. In the Gold-Hoyle model, the azimuthal and axial field components are given as:

\begin{subequations}
  \begin{equation}
    \label{eq-3a}
      B_{\phi} = \dfrac{B_0 \tau r}{1 + \tau^2r^2},
  \end{equation}
  \begin{equation}
    \label{eq-3b}
    B_z = \dfrac{B_0}{1 + \tau^2 r^2},
  \end{equation}
\end{subequations}

\noindent where $B_0$ is the axial field strength, $r$ is the radial distance from the rope axis, and $\tau$ is the angle a field line rotates about the axis from the leading edge of the flux rope to the trailing edge. The Gold-Hoyle rope is therefore very different to the Lundquist rope, in which the field-line twist is at a minimum at the rope axis and infinite at the rope boundaries. 

\begin{figure*}
\centering
\includegraphics[width=\textwidth]{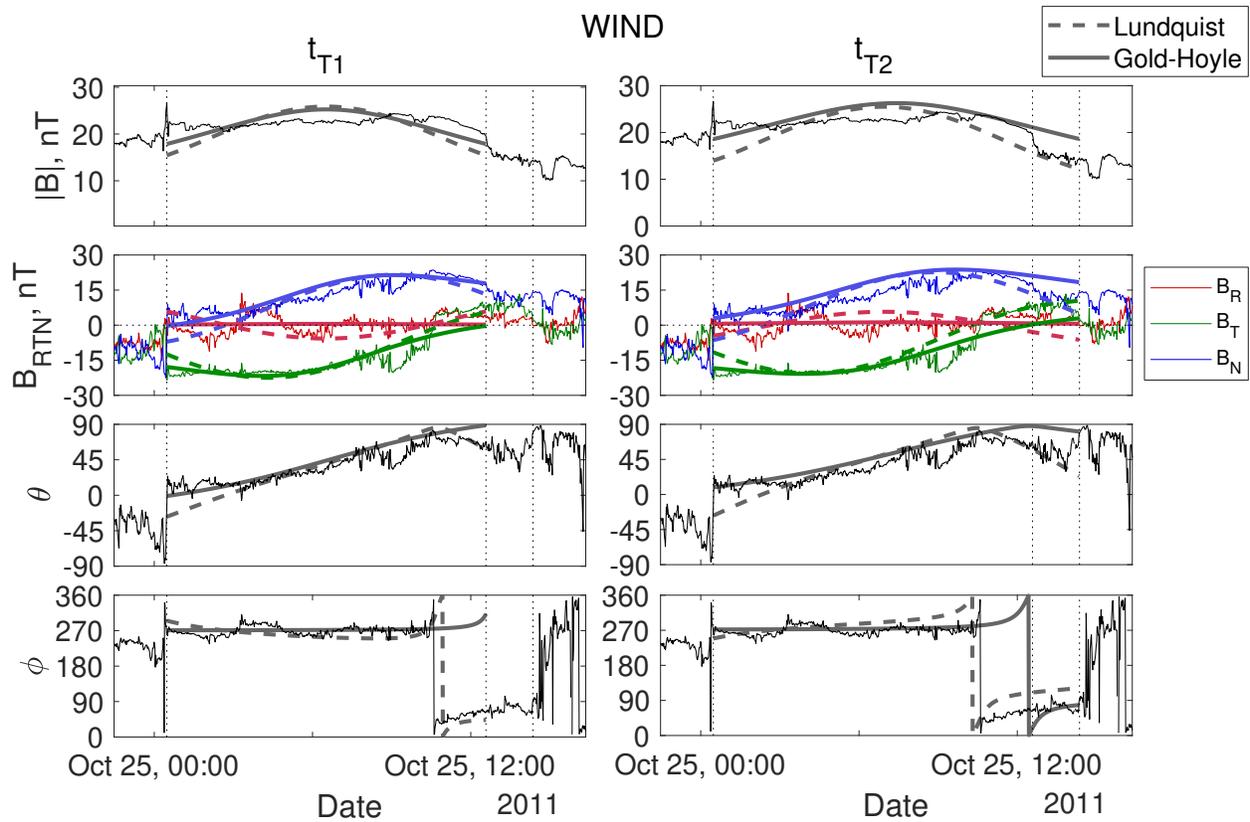}
\captionsetup{font=small, labelfont=bf}
\caption{Force-free flux rope models fitted to the in situ magnetic field signatures observed by Wind, displayed in a similar format as Figure \ref{fig:wind}. The Lundquist model (dashed line) and Gold-Hoyle model (solid line) fitting is shown using both the earlier trailing edge (left-hand side) and the later trailing edge (right-hand side).}
\label{fig:wind_fits}
\end{figure*}

\begin{figure*}
\centering
\includegraphics[width=\textwidth]{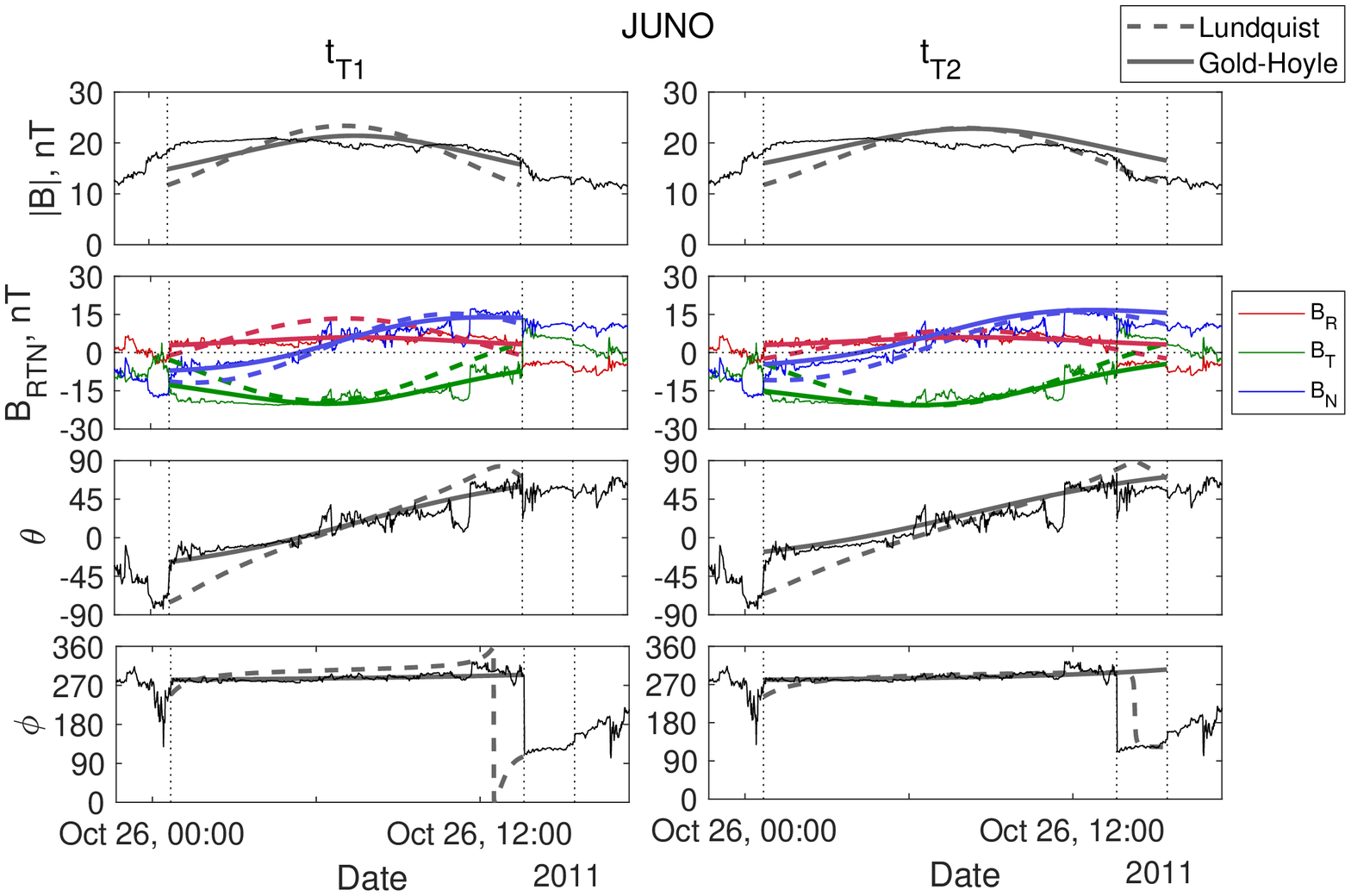}
\captionsetup{font=small, labelfont=bf}
\caption{Force-free flux rope models fitted to the in situ magnetic field signatures observed by Juno, displayed in the same format as Figure \ref{fig:wind_fits}.}
\label{fig:juno_fits}
\end{figure*}

Figures \ref{fig:wind_fits} and \ref{fig:juno_fits} present the Lundquist (dashed line) and Gold-Hoyle (solid line) fits to the magnetic field data of the flux rope using both the earlier trailing edge boundary (left-hand side) and the later trailing edge (right-hand side) at Wind and Juno, respectively. The fitting at ACE, THEMIS B and C is presented as electronic supplementary material (Figures 10, 11, and 12). Visual inspection of Figure \ref{fig:wind_fits} shows that both models fit the magnetic field data to a good approximation. The main difference between the two models can be seen to be in how they represent the weakest radial component. The goodness of fit of the Lundquist model fits at Wind is marginally better than the Gold-Hoyle model using the earlier trailing edge, but is more similar using the later trailing edge. At Juno, Figure \ref{fig:juno_fits} shows that the Lundquist and Gold-Hoyle model  fits are of a comparatively similar goodness using the later trailing edge, similar to those at Wind. At Juno, the largest difference in the fits arises when using the earlier trailing edge boundary -  we find that the Gold-Hoyle model fits much better to the magnetic field data than the Lundquist model using this boundary. Looking at the large scale rope structure, inspection of the model fits shows that the two models follow similar patterns across both spacecraft and both trailing edges. The exception to this pattern is seen in the Lundquist fitting at Wind where the weakest radial component of the field is of opposite sign for the different trailing edge times.

Both force-free fitting methods allow for estimates to be made of various global cloud properties such as the axial field strength, $B_0$, its radial width, $D'$, the normalised closest approach distance of the spacecraft to the flux rope axis known as the impact parameter, $p$, and the flux rope axis orientation.

\begin{figure*}
\centering
\includegraphics[width=\textwidth]{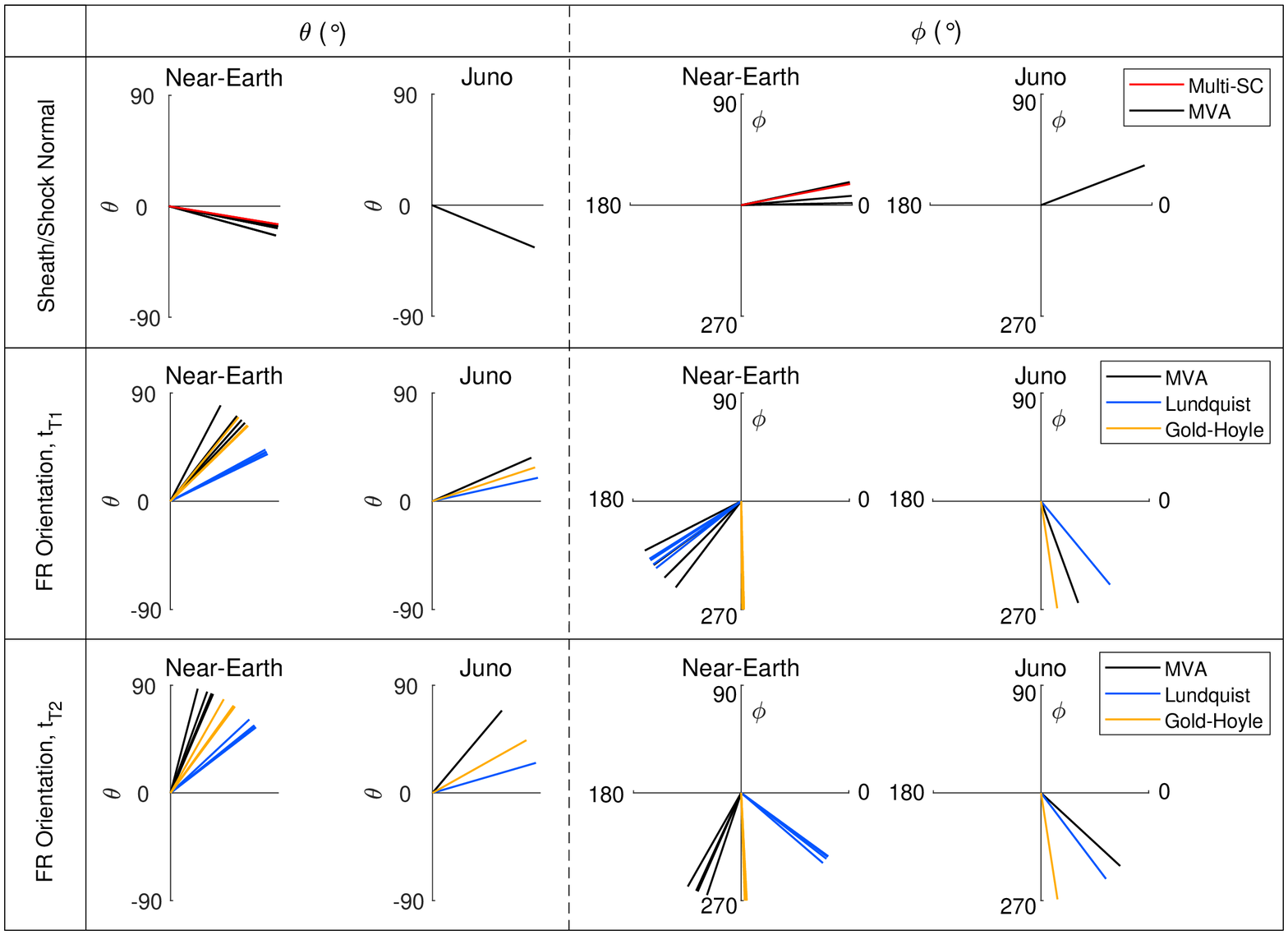}
\captionsetup{font=small, labelfont=bf}
\caption{Calculated orientations presented in RTN coordinates, where $\theta$ is the angle to the R-T plane, and $\phi$ is the angle swept out anticlockwise from the Sun-Earth line, projected onto the R-T plane ($\phi$ = 180$^\circ$ points Sunwards). $\theta$ is presented for the near-Earth spacecraft and Juno in the first column, and $\phi$ is presented for the near-Earth spacecraft and Juno in the second column. The normal directions to the sheath calculated using MVA (black) and the propagation direction of the shock calculated using timing considerations between the near-Earth spacecraft (red) are presented in the first row. The flux rope orientations calculated by MVA (black), Lundquist (blue), and Gold-Hoyle fitting (orange) using the earlier trailing edge are presented in the second row, and in the third row using the later trailing edge.}
\label{fig:orientations}
\end{figure*}

Table 2 also summarises the results of the force-free fitting. Both the Lundquist and Gold-Hoyle fitting methods show that the rope is consistently left-handed ($H=-1$) at the near-Earth spacecraft and Juno. Figure \ref{fig:orientations} presents the flux rope orientations calculated using the earlier trailing edge (second row) and the later trailing edge (third row). The Lundquist fitting gives a mean flux rope orientation of $\theta = 27.1/39.5^{\circ}$ and $\phi = 214.8/322.3^{\circ}$ at the near-Earth spacecraft and an orientation of $\theta = 12.5/16.2^{\circ}$ and $\phi = 309.6/307.1^{\circ}$ at Juno. The Gold-Hoyle fitting gives a mean orientation of $\theta = 45.9/55.4^{\circ}$ and $\phi = 271.1/272.5^{\circ}$ at the near-Earth spacecraft and an orientation of $\theta =  18.2/29.3^{\circ}$ and $\phi = 278.6/278.8^{\circ}$ at Juno. The Lundquist and Gold-Hoyle fits at each spacecraft meet the $\chi^{2}$ and $\delta$ error requirements for reasonably accurate fits \citep[see][for further details on these parameters]{good2019self}, i.e. they fit well to the data. For such fits, \citet{lepping2003estimated} estimated errors of $\approx 13^{\circ}$ and $\approx 30^{\circ}$ in $\theta$ and $\phi$, respectively. The $\chi^{2}$ values are listed in Table 2 and show comparable trends to those observed by visual inspection of the fitting: the Lundquist model fits at Juno are comparatively worse than at Wind using the later trailing edge but are of similar goodness of fit using the earlier trailing edge, whereas the Gold-Hoyle model fits at Juno are comparatively better than at Wind using the earlier trailing edge, but similar using the later trailing edge. The force-free fitting orientations at the near-Earth spacecraft are also in reasonable agreement with those found in previous studies for the same ICME that used Wind data; \citet{lepping2015wind} found that $\theta = 40^{\circ}$ and $\phi = 291^{\circ}$ and \citet{wood2017stereo} found that $\theta = 45^{\circ}$ and $\phi = 286^{\circ}$. The flux rope orientations and left-handedness are supported by the flux rope classifications observed in Section \ref{sec:scobservations} (ENW at the near-Earth spacecraft and SEN at Juno). 

\begin{landscape}
\vspace*{\fill}
\noindent\begin{minipage}{\linewidth}
\centering
\captionsetup{font=small, labelfont=bf}
\captionof{table}{The results of the analysis performed at each spacecraft, organised by method. Multi-spacecraft timing considerations at the near-Earth spacecraft determine a propagation direction of the ICME shock front. MVA has been applied to both the sheath to determine the normal direction to the sheath, and the flux rope to determine the flux rope axis orientation. The flux rope axis orientation is also determined by the force-free fitting methods (Lundquist and Gold-Hoyle). Both force-free flux rope fitting methods give estimates of the axial magnetic field strength, $B_0$, handedness, $H$, impact parameter, $p$, minimised chi-squared, $\chi^2$, and modelled radial width, $D'$.}
\resizebox{\linewidth}{!}{\begin{tabular}{c|c|ccccc}
\hline\hline
\begin{tabular}[c]{@{}c@{}}Analysis \\ Method \end{tabular} & Parameter & Wind  & ACE  & THEMIS B  & THEMIS C  & Juno  \\ 
\hline
\multirow{2}{*}{Multi-S/C} & \multirow{2}{*}{Shock propagation direction} & \multicolumn{4}{c}{$\theta_p$ = -9.4$^\circ$  } & - \\
 &  & \multicolumn{4}{c}{$\phi_p$ = 11.0$^\circ$  } & - \\ 
\hline
\multirow{5}{*}{MVA} & \multirow{2}{*}{Sheath normal direction} & $\theta_s$ = -11.4$^\circ$  & $\theta_s$ = -15.3$^\circ$   & $\theta_s$ = -10.2$^\circ$   & $\theta_s$ = -10.5$^\circ$   & $\theta_s$ = -22.4$^\circ$   \\
 &  & $\phi_s$ = 4.8$^\circ$   & $\phi_s$ = 1.1$^\circ$   & $\phi_s$ = 11.6$^\circ$   & $\phi_s$ = 12.0$^\circ$   & $\phi_s$ = 21.0$^\circ$   \\
 & \multirow{2}{*}{FR axis orientation} & $\theta_a$ = 46.4$^\circ$ /70.1$^\circ$   & $\theta_a$ = 62.3$^\circ$ /75.4$^\circ$   & $\theta_a$ = 48.7$^\circ$ /66.7$^\circ$   & $\theta_a$ = 52.0$^\circ$ /67.6$^\circ$   & $\theta_a$ = 23.7$^\circ$ /49.9$^\circ$   \\
 &  & $\phi_a$ = 207.1$^\circ$/ 251.5$^\circ$   & $\phi_a$ = 232.8$^\circ$ /245.3$^\circ$   & $\phi_a$ = 216.1$^\circ$ /240.3$^\circ$   & $\phi_a$ = 224.9$^\circ$ /246.2$^\circ$   & $\phi_a$ = 290.1$^\circ$ /317.3$^\circ$   \\
 & FR eigenvalue ratios, $\dfrac{\lambda_1}{\lambda_2}$, $\dfrac{\lambda_3}{\lambda_2}$  & 7.56/6.36, 0.61/0.46 & 7.88/6.17, 0.42/0.26 & 7.92/6.74, 0.48/0.37 & 7.51/6.24, 0.48/0.34 & 13.80/3.68, 0.31/0.05 \\
\hline
\multirow{6}{*}{Lundquist} & \multirow{2}{*}{FR axis orientation} & $\theta_a$ = 25.8$^\circ$ /38.9$^\circ$   & $\theta_a$ = 28.5$^\circ$ /43.0$^\circ$   & $\theta_a$ = 26.9$^\circ$ /37.5$^\circ$   & $\theta_a$ = 27.3$^\circ$ /38.5$^\circ$   & $\theta_a$ = 12.5$^\circ$ /16.2$^\circ$   \\
 &  & $\phi_a$ = 213.1$^\circ$ /322.3$^\circ$   & $\phi_a$ = 215.7$^\circ$ /324.0$^\circ$   & $\phi_a$ = 212.1$^\circ$ /323.5$^\circ$   & $\phi_a$ = 218.2$^\circ$ /319.4$^\circ$   & $\phi_a$ = 309.6$^\circ$ /307.1$^\circ$   \\
 & Axial field strength, $B_0$ [nT]  & 30.8/27.9 & 30.8/28.3 & 32.0/29.0 & 30.8/28.3 & 23.4/23.5 \\
 & Handedness, $H$  & -1/-1 & -1/-1 & -1/-1 & -1/-1 & -1/-1 \\
 & Impact parameter, $p$  & 0.476/0.341 & 0.475/0.346 & 0.477/0.341 & 0.452/0.322 & 0.047/0.181 \\
 & Minimised chi-squared, $\chi^2$  & 0.125/0.124 & 0.089/0.108 & 0.131/0.122 & 0.113/0.114 & 0.234/0.120 \\
 & Modelled radial width, $D'$ [AU]  & 0.102/0.131 & 0.107/0.131 & 0.100/0.121 & 0.104/0.125 & 0.115/0.136 \\
\hline
\multirow{6}{*}{\begin{tabular}[c]{@{}c@{}}Gold-\\ Hoyle \end{tabular}} & \multirow{2}{*}{FR axis orientation} & $\theta_a$ = 44.1$^\circ$ /54.2$^\circ$   & $\theta_a$ = 50.9$^\circ$ /60.4$^\circ$   & $\theta_a$ = 44.6$^\circ$ /53.8$^\circ$   & $\theta_a$ = 44.1$^\circ$ /53.3$^\circ$   & $\theta_a$ = 18.2$^\circ$ /29.3$^\circ$   \\
 &  & $\phi_a$ = 270.9$^\circ$ /272.3$^\circ$   & $\phi_a$ = 270.5$^\circ$ /271.9$^\circ$   & $\phi_a$ = 271.5$^\circ$ /273.0$^\circ$   & $\phi_a$ = 271.5$^\circ$ /272.9$^\circ$   & $\phi_a$ = 278.6$^\circ$ /278.8$^\circ$   \\
 & Axial field strength, $B_0$ [nT]  & 25.3/26.3 & 25.8/26.8 & 26.2/27.2 & 25.8/26.8 & 21.6/23.0 \\
 & Handedness, $H$  & -1/-1 & -1/-1 & -1/-1 & -1/-1 & -1/-1 \\
 & Impact parameter, $p$  & 0.011/0.023 & 0.006/0.016 & 0.019/0.043 & 0.019/0.030 & 0.153/0.139 \\
 & Minimised chi-squared, $\chi^2$  & 0.089/0.093 & 0.081/0.085 & 0.084/0.086 & 0.083/0.086 & 0.040/0.149 \\
 & Modelled radial width, $D'$ [AU] & 0.137/0.156 & 0.135/0.152 & 0.132/0.148 & 0.130/0.147 & 0.148/0.164 \\
\hline
\end{tabular}
}
\end{minipage}
\vspace*{\fill}
\label{tab:analysis}
\end{landscape}

Comparing the flux rope axis orientations for each of the force-free flux rope fitting methods and those calculated using MVA, we find: 

\begin{enumerate}[label=\roman*., itemsep=0pt, topsep=0pt]
    \item  MVA results in the highest inclination to the SE plane for both flux rope trailing edges, with a mean $\theta = 52.4/70.0^{\circ}$ for the near-Earth spacecraft. This is in comparison to $\theta = 27.1/39.5^{\circ}$ for the Lundquist model, and $\theta = 45.9/55.4^{\circ}$ for the Gold-Hoyle model. Comparing $\phi$ angles at the near-Earth spacecraft, we find that the mean $\phi = 220.2/245.8^{\circ}$ for MVA, $\phi = 214.8/322.3^{\circ}$ for the Lundquist model, and $\phi = 271.1/272.5^{\circ}$ for the Gold-Hoyle model. Note that the Lundquist model results in the highest and lowest mean value of $\phi$ given the different trailing edge times.
    \item Comparing the results for each method at the near-Earth spacecraft to Juno, we find that the flux rope orientation differs by $\theta = 28.7/20.1^{\circ}$, $14.6/23.3^{\circ}$, and $27.7/26.1^{\circ}$ towards the SE plane, and $\phi = 69.9/71.5^{\circ}$, $94.8/-15.2^{\circ}$, $7.5/6.3^{\circ}$ anticlockwise in the SE plane, for MVA, Lundquist and Gold-Hoyle models, respectively. The overall trend in flux rope orientation between the near-Earth spacecraft and Juno is demonstrated clearly in Figure \ref{fig:orientations} by comparing the subplots of each panel: the orientation is closer to the SE plane at Juno, and rotates anticlockwise between the near-Earth spacecraft and Juno. The difference in $\theta$ exceeds the $\approx 13^{\circ}$ uncertainty found by \citet{lepping2003estimated} in the orientation for all methods/trailing edges. However, this is not the case for each $\phi$ angle where just MVA and the earlier Lundquist value exceed the $\approx 30^{\circ}$ uncertainty. The difference in orientation towards the SE plane is quite significant over a relatively small radial separation. As discussed in Section \ref{sec:context}, the source of the faster solar wind following the event is unclear, but perhaps may have had an effect on the change in flux rope orientation between the near-Earth spacecraft and Juno. Different parts of an ICME can also evolve in a different manner in the structured solar wind which could lead to differences in flux rope properties, such as orientation, at spacecraft separated in longitude \citep[e.g.][]{savani2010,owens2017}. However, over the small longitudinal and radial separation in this case, such differences would not be expected to be large.
    \item Changing between the two trailing edge times (comparing the second and third rows of Figure \ref{fig:orientations}) produces the largest difference in $\theta$ with MVA - an average difference of 17.6$^{\circ}$, compared to 12.4$^{\circ}$ and 9.5$^{\circ}$ using the Lundquist and Gold-Hoyle models, respectively. However, the largest difference in the $\phi$ angle is produced by the Lundquist model - 107.5$^{\circ}$ compared to 25.6$^{\circ}$ for MVA, and 1.4$^{\circ}$ for the Gold-Hoyle model. For these cases, the Gold-Hoyle model is least affected by the change in trailing edge time defined for the flux rope, whereas the Lundquist model is more sensitive to this, especially in the resulting fit to the radial component of the magnetic field. A previous study by \citet{demoulin2018exploring} explores the sensitivity of flux rope orientation with changing flux rope boundaries for MVA and finds that similarly, the boundaries defined strongly affect the resulting flux rope orientation. 
    \item Considering the uncertainty in the orientations at the near-Earth spacecraft, the Lundquist model is the least variant in results for $\theta$, with a standard deviation of 1.1/2.4$^{\circ}$ in comparison to 7.0/3.9$^{\circ}$ for MVA, and 3.3/3.3$^{\circ}$ for Gold-Hoyle. However, the Gold-Hoyle model is the least variant in results for $\phi$, with a standard deviation of 0.5/0.5$^{\circ}$, compared with 11.1/4.6$^{\circ}$ for MVA, and 2.7/2.1$^{\circ}$ for Lundquist. This is clearly demonstrated by the spread of orientations in Figure \ref{fig:orientations}. Overall, MVA produces the widest spread in results for the near-Earth spacecraft, with the force-free fitting models performing similarly.
    \item The difference between mean results at the near-Earth spacecraft across methods shows that for $\theta$, the Gold-Hoyle model results are most similar to MVA, with a mean difference of $6.4/14.5^{\circ}$. This is in comparison with Lundquist and MVA with a mean difference of $25.2/30.5^{\circ}$, and Lundquist and Gold-Hoyle with a mean difference of $18.8/16.0^{\circ}$. The $\phi$ angle is more dependent on the trailing edge time, where the difference between MVA and Lundquist is $5.5/76.5^{\circ}$, MVA and Gold-Hoyle is $50.9/26.7^{\circ}$, and Lundquist and Gold-Hoyle is $56.3/49.8^{\circ}$. 
\end{enumerate}

\vspace{0.2cm}
The closer to the axis a spacecraft crosses the flux rope, the more reliable the estimated MVA axis orientation and calculated force-free fitting parameters have been found to be \citep{klein1982interplanetary, bothmer1997structure, farrugia1999uniform, xiao2004inferring, gulisano2005magnetic, gulisano2007estimation, ruffenach2012multispacecraft, ruffenach2015statistical}. Investigating the impact parameters at each spacecraft, we find that the Lundquist model suggests that Juno passes through the flux rope closer to the central axis than any of the near-Earth spacecraft (at Juno, $p = 0.047/0.181$, in comparison to the near-Earth spacecraft where the mean $p = 0.470/0.338$), whereas the Gold-Hoyle model suggests that the near-Earth spacecraft pass closest to the flux rope axis ($p =  0.153/0.139$ at Juno and the mean $p = 0.014/0.028$ at the near-Earth spacecraft). Comparing fitting methods, the impact parameters are most similar for Juno, with more of a contrast between values for the near-Earth spacecraft. The two fitting methods give very different impact parameters due to the magnetic field geometry: the Lundquist $p$ values are higher than the Gold-Hoyle values because an intermediate-$p$ Lundquist flux rope encounter, in which the field is not observed to rotate by a full 180$^{\circ}$ between leading and trailing edges, is similar to a low-$p$ Gold-Hoyle rope encounter. 

If we consider the propagation of a perfectly cylindrical flux rope with a concentric sheath region draped ahead of the flux rope, the sheath normal and the flux rope axis should ideally be perpendicular to each other at each spacecraft. Comparing the flux rope axis orientations of each method/defined trailing edge to that of the normal direction to the sheath, we find the mean angle between these vectors for the near-Earth spacecraft as $\delta = 48.0/68.2^{\circ}$, $\delta = 30.1/66.6^{\circ}$, and $\delta = 77.0/77.3^{\circ}$, with the MVA, Lundquist, and Gold-Hoyle methods, respectively. This angle can be visualised by comparing the second and third rows of Figure \ref{fig:orientations} to the first. For the near-Earth spacecraft, we therefore find that the Gold-Hoyle model produced values closer to the ideal, and across all methods, using the flux rope axis from fits with the later trailing edge produced results consistently closer to the ideal than those using the earlier trailing edge. At Juno, $\delta = 80.4/88.5^{\circ}$, $\delta = 78.2/82.0^{\circ}$, and $\delta = 72.0/69.0^{\circ}$ for the MVA, Lundquist, and Gold-Hoyle methods, respectively. Overall, the values are closer to the ideal at Juno for each method/trailing edge time except for the Gold-Hoyle model.

The axial field strength estimated by the Lundquist model is consistently higher than the corresponding value estimated by the Gold-Hoyle model: the Lundquist fitting gives the mean $B_0 = 31.1/28.4$ nT at the near-Earth spacecraft and $B_0 = 23.4/23.5$ nT at Juno, and the Gold-Hoyle fitting gives the mean $B_0 = 25.8/26.8$ nT at the near-Earth spacecraft and $B_0 = 21.6/23.0$ nT at Juno. Investigating the relationship between the axial field strength given by the force-free fitting models and the heliocentric distance at which it was measured, we find a power law for the Lundquist results that suggests $B_0 \propto {r_H}^{-1.25 \pm 0.03}/r_H^{-0.83 \pm 0.03}$, and for the Gold-Hoyle results we find $B_0 \propto r_H^{-0.77 \pm 0.04}/r_H^{-0.67 \pm 0.04}$. The earlier and later trailing edge Gold-Hoyle fit relationships and the later trailing edge Lundquist fit relationship are similar to the relationships derived using in situ observations in Section \ref{sec:scobservations}. Interestingly, the relationship derived using the earlier trailing edge Lundquist fits is consistent with the previously mentioned relationships derived by \citet{richardson2014identification} and \citet{ebert2009bulk} as the radial dependence may be quite different for $B_0$ derived from fits in comparison to the field parameters observed in situ. However, as previously discussed, disagreement between the relationships derived in this study with previous studies is likely due to the small radial separation of 0.24 AU between spacecraft observations, and therefore the difference in longitude between observations and thus the different path taken by the spacecraft through the flux rope is likely to be the dominant effect. 

The observed radial width ($D$) does not take into account the orientation of the flux rope as it passes the spacecraft, nor the impact parameter. To correct for this, a scale factor, estimated from the force-free fitting may be applied to the observed radial width to give an estimate of the true flux rope width, $D'$. Hence $D' = S D = S v_c (t_T - t_L)$, where $S$ is the scale factor that can be less than, equal to, or greater than 1 and accounts for both the impact parameter and the rope orientation. The calculated modelled radial widths are listed in Table 2. For the Lundquist model, a scaling value less than 1 is given to correct for the orientation of the flux rope at each spacecraft. The Lundquist fitting gives the mean $D' = 0.103/0.127$ AU at the near-Earth spacecraft, and $D' = 0.115/0.136$ AU at Juno. The Gold-Hoyle fitting only produces a non-unity scale factor for Juno, however, these values are still very close to unity resulting in only a 0.001 AU difference to the radial width observed for the earlier trailing edge: $D' = 0.148/0.164$ AU, in comparison to $D =  0.147/0.164$ AU. The mean calculated radial width remains the same as the observed radial width at the near-Earth spacecraft, $D' = D = 0.134/0.151$ AU, likely due to the flux rope orientation lying close to perpendicular with the radial direction ($\phi$= 271.1/272.5$^\circ$) and the small impact parameter, and therefore little adjustment is necessary for the spacecraft path length through the flux rope. The modelled radial widths are similarly consistent between the near-Earth spacecraft, and show that there was very little expansion over the short distance between Earth and Juno for both models and trailing edges. The difference in modelled radial width between Wind and Juno is 0.013/0.005 AU for the Lundquist model and 0.011/0.008 for Gold-Hoyle. These values are consistent with the difference in observed radial width, and are less than the expected 0.019/0.024 AU calculated in Section \ref{sec:scobservations}.  

\section{Summary and Conclusions} \label{sec:conclusion}

An ICME that caused a strong geomagnetic storm at Earth commencing on 24 October 2011 was observed in situ by ACE, Wind, ARTEMIS, and Juno. The geomagnetic storm was the strongest recorded in 2011, with a minimum Dst of -147 nT at its peak. The ICME displayed a clear magnetic flux rope structure which has been analysed using two fitting models, Lundquist and Gold-Hoyle, and MVA. 

Due to the positioning of the spacecraft in the near-Earth environment, ACE, Wind, and the two ARTEMIS spacecraft, THEMIS B and THEMIS C, have been used to perform multi-spacecraft analysis in conjunction with Juno which had recently commenced its cruise phase to Jupiter and was therefore close to radial alignment with the near-Earth spacecraft, longitudinally separated by only 3.6$^{\circ}$ throughout the event. During this time, the radial separation between Juno and the near-Earth spacecraft was 0.24 AU. Cases where spacecraft are separated by such radial distances and longitudinal separations are rare, and therefore these observations have allowed for an interesting analysis of the evolution of a magnetic cloud and evaluation of whether radial or longitudinal effects dominate.

We find that the overall magnetic field magnitude profiles, as well as the behaviour of the magnetic field components, are similar between the investigated spacecraft. However, we have also found that observations made in situ between the near-Earth spacecraft and Juno display some significant differences, despite the small longitudinal separation between the spacecraft; e.g. we observe a sudden increase in radial speed of the solar wind following the flux rope at the near-Earth spacecraft but not at Juno. These differences can arise from evolution in time and/or from longitudinal separation, although Juno and Earth are relatively close to each other both radially and longitudinally. Significant differences have previously also been reported in ICME flux rope properties over relatively small longitudinal separations of only a few degrees \citep{kilpua2011multipoint, winslow2016longitudinal}. \citet{lugaz2018} recently reported considerable differences in the magnetic field components for a magnetic cloud they observed near the Earth orbit where spacecraft were only 0.01 AU apart. Studies based on more widely separated (several degrees in longitude) spacecraft have reported highly different flux rope orientations, suggesting this to be a local rather than global parameter \citep[e.g.][]{savani2010,farrugia2011,mostl2012multi}. 

The flux rope orientation differs between the near-Earth spacecraft and Juno by $\theta = 28.7/20.1^{\circ}$, $14.6/23.3^{\circ}$, and $27.7/26.1^{\circ}$ (dependent on earlier/later trailing edge) towards the SE plane, and $\phi = 69.9/71.5^{\circ}$, $94.8/-15.2^{\circ}$, $7.5/6.3^{\circ}$ anticlockwise in the SE plane, for MVA, Lundquist and Gold-Hoyle models, respectively. The orientation of the flux rope axis has been shown to have a clear difference in $\theta$ despite the relatively small spacecraft separations, irrespective of analysis method or trailing edge defined. However, the difference in $\phi$ only exceeds uncertainties in orientation for the MVA values and the earlier trailing edge value for the Lundquist method. We propose that the difference in flux rope orientation is not necessarily just due to the radial evolution of the ICME, but more so due to the longitudinal separation of the spacecraft, despite this being small. This ambiguity, inherent to the localised nature of in situ measurements, has been a clear feature of previous alignment studies up to 1 AU; we note in agreement with a more limited number of previous studies \citep[e.g.][]{mulligan1999intercomparison} that this ambiguity must also be taken into account when analysing ICMEs beyond 1 AU. \citet{winslow2016longitudinal} found that in situ observations of an ICME between the MErcury Surface, Space ENvironment, GEochemistry, and Ranging (MESSENGER) spacecraft and STEREO-A were significantly affected due to interactions between the ICME and a heliospheric plasma sheet/current sheet, despite a small longitudinal separation of just 3$^{\circ}$. Similarly, in this study we note that differences between in situ observations may have arisen due to the sudden increase in solar wind speed following the flux rope observed at the near-Earth spacecraft but not at Juno, the source of which remains unclear.

Comparing the force-free fitting models, both the Lundquist and Gold-Hoyle methods give broadly similar axis directions that are consistent with the ENW/SEN flux rope classifications. At the near-Earth spacecraft, there is a mean difference of $\theta = 18.8/16.0^{\circ}$ and $\phi = 56.3/49.8^{\circ}$ between the fitting methods. The Lundquist model is least variant in $\theta$, with a standard deviation of $1.1/2.4^{\circ}$ in comparison to $3.3/3.3^{\circ}$ for Gold-Hoyle. However, the Gold-Hoyle model is the least variant in $\phi$, with a standard deviation of $0.5/0.5^{\circ}$, compared with $2.7/2.1^{\circ}$ for Lundquist. Despite the difference in results, the similar standard deviations and visual inspection of the fitting at the near-Earth spacecraft show that overall, both force-free fitting models performed similarly, and give comparatively good fits to the data.

As discussed in Sections \ref{sec:scobservations} and \ref{sec:analysis}, relationships found between the observed mean and maximum field strengths and axial field strengths given by the force-free fitting with heliocentric distance were mostly in disagreement with the relationships found in previous studies at distances greater than 1 AU \citep{ebert2009bulk, richardson2014identification}. The disagreement is likely the result of using observations/parameters calculated for five spacecraft over a relatively short radial separation of 0.24 AU, whereas the previous studies used a large number of events observed between 1 and 5.4 AU, thus differences in magnetic field due to the small longitudinal separation dominate.

In conclusion, this case study demonstrates that Juno cruise data is a potentially valuable resource for studies, including multi-spacecraft studies, of the evolution of ICME magnetic fields between 1 and 5 AU, and further demonstrates that caution should be exercised in radial alignment studies. The presence of increased solar wind speed following the event at Wind but not at Juno shows that even small longitudinal separations of a few degrees between spacecraft can still result in significantly different observations and event properties.

\newpage
\footnotesize{\textbf{Acknowledgements} We have benefited from the availability of ACE, Wind, ARTEMIS, Juno, SDO, and STEREO data, and thus would like to thank the instrument teams and the PDS:PPI and SPDF CDAWeb data archives for their distribution of data. We are very grateful to the referee for the insightful and constructive comments that helped to improve the manuscript. E.D. would also like to thank Lucie Green and Mathew Owens for the useful discussions regarding the solar sources and background solar wind, respectively. This research was supported by funding from the STFC studentship ST/N504336/1 (E.D.). The work of E.K. and S.G.  has received funding from the European Research Council (ERC) under the European Union's Horizon 2020 research and innovation programme (ERC-COG 724391). E.K. and S.G. acknowledges Academy of Finland project SMASH no. 310445. The results of E.K. and S.G. presented here have been achieved under the framework of the Finnish Centre of Excellence in Research 5 of Sustainable Space (FORESAIL; Academy of Finland grant numbers 312390), which we gratefully acknowledge.} 

\section*{Disclosure of Potential Conflicts of Interest}

The authors declare that there are no conflicts of interest.

\bibliographystyle{spr-mp-sola}
{\small\bibliography{solar-bib.bib}}

\section*{Electronic Supplementary Material}

\begin{figure*}
\centering
\captionsetup{font=small, labelfont=bf}
\includegraphics[width=\textwidth]{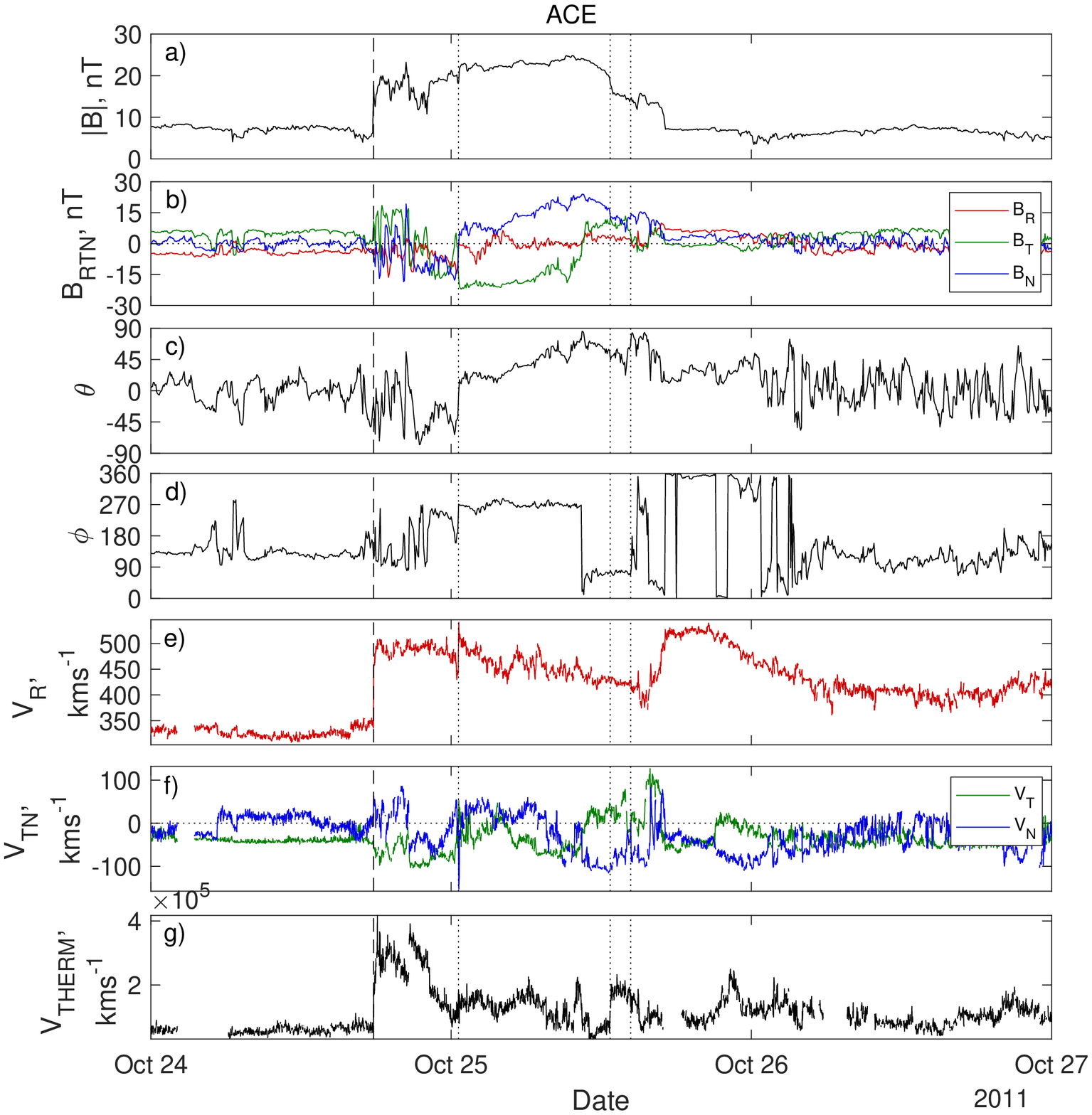}
\caption{In situ magnetic field (1 minute resolution) signatures observed by ACE displayed in the same format as Figure \ref{fig:wind}.}
\label{fig:ace_obs}
\end{figure*}

\begin{figure*}
\centering
\captionsetup{font=small, labelfont=bf}
\includegraphics[width=\textwidth]{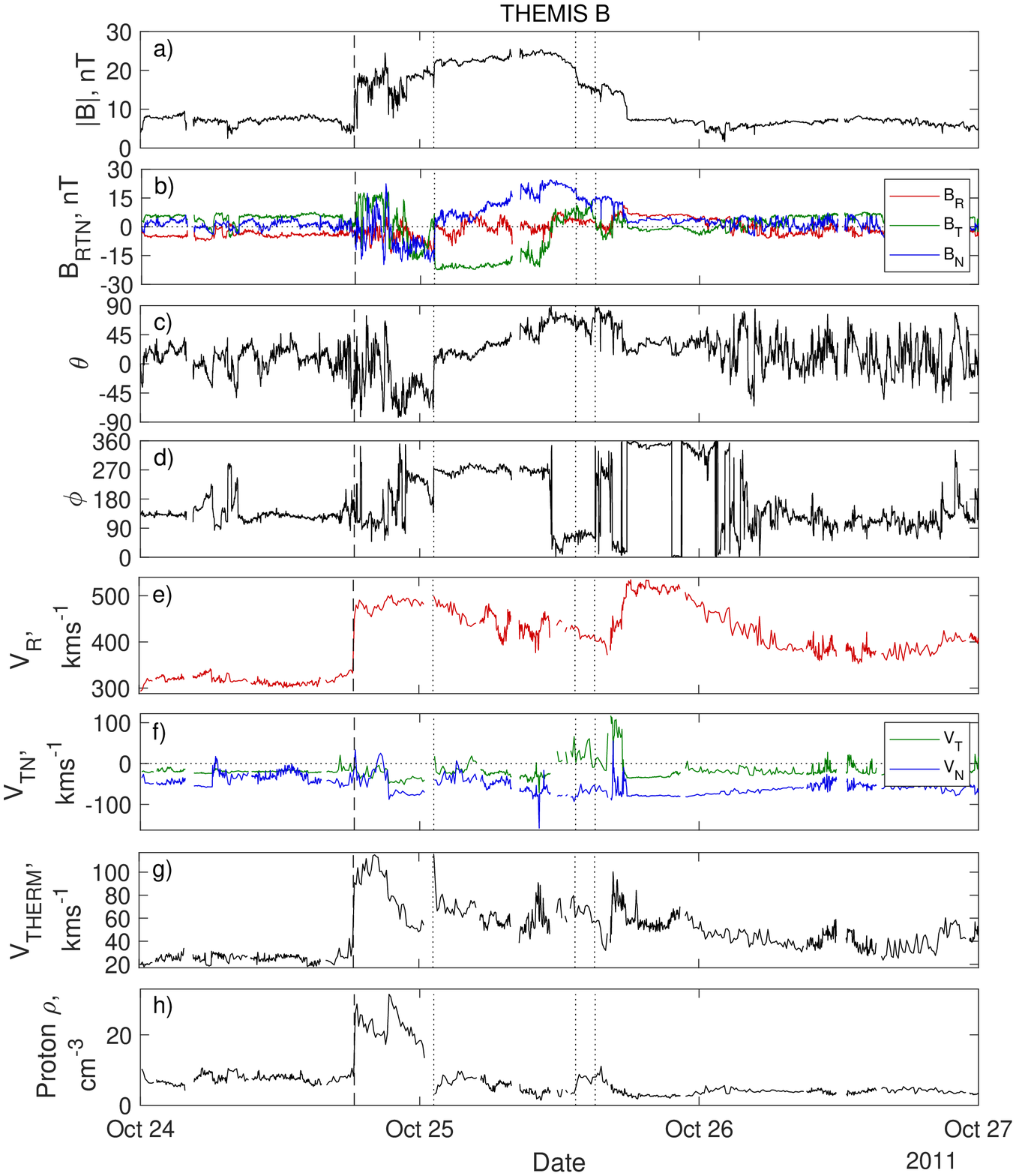}
\caption{In situ magnetic field (1 minute resolution) signatures observed by THEMIS B displayed in the same format as Figure \ref{fig:wind}.}
\label{fig:thb_obs}
\end{figure*}

\begin{figure*}
\centering
\captionsetup{font=small, labelfont=bf}
\includegraphics[width=\textwidth]{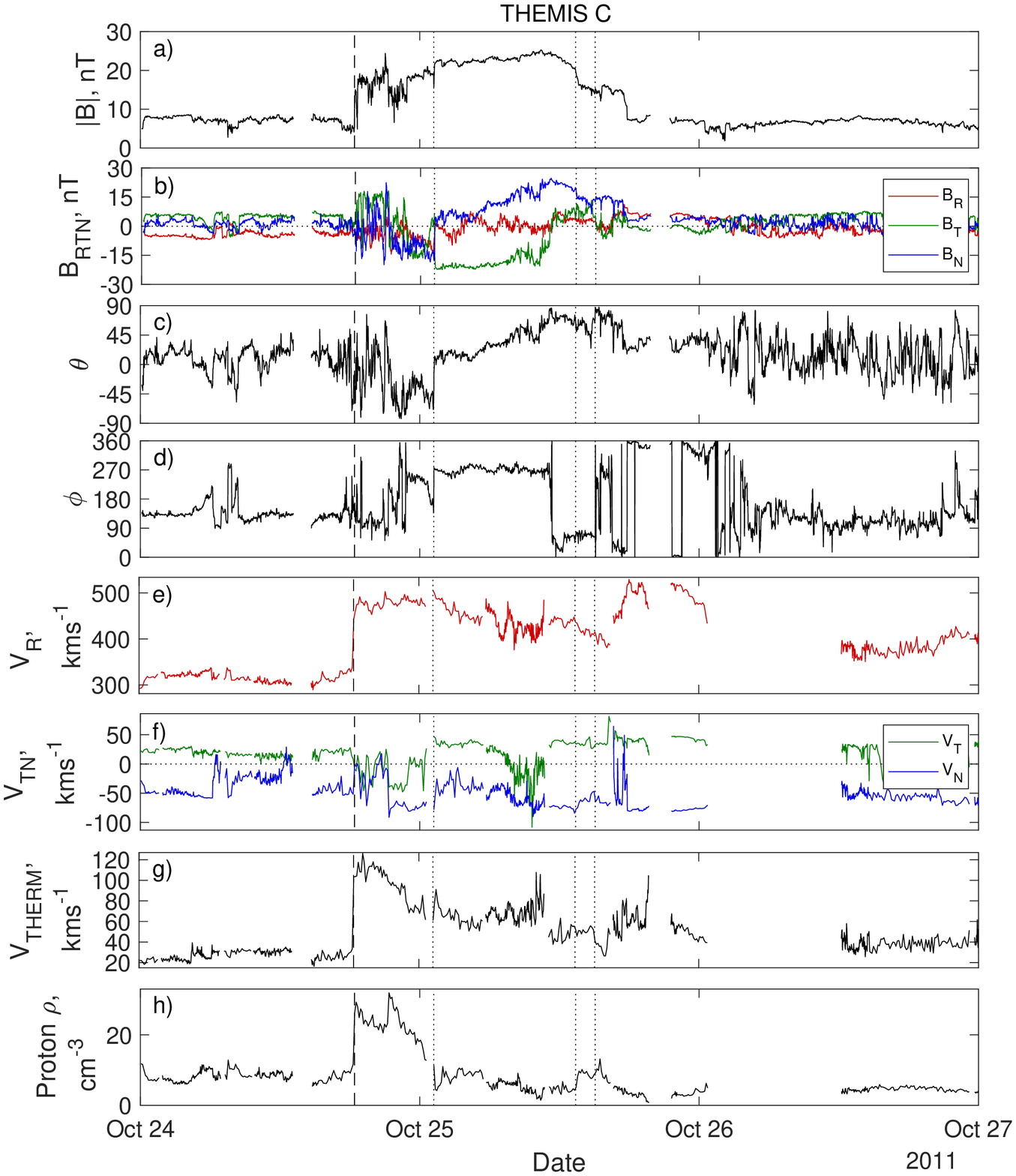}
\caption{In situ magnetic field (1 minute resolution) signatures observed by THEMIS C displayed in the same format as Figure \ref{fig:wind}.}
\label{fig:thc_obs}
\end{figure*}

\begin{figure*}
\centering
\captionsetup{font=small, labelfont=bf}
\includegraphics[width=\textwidth]{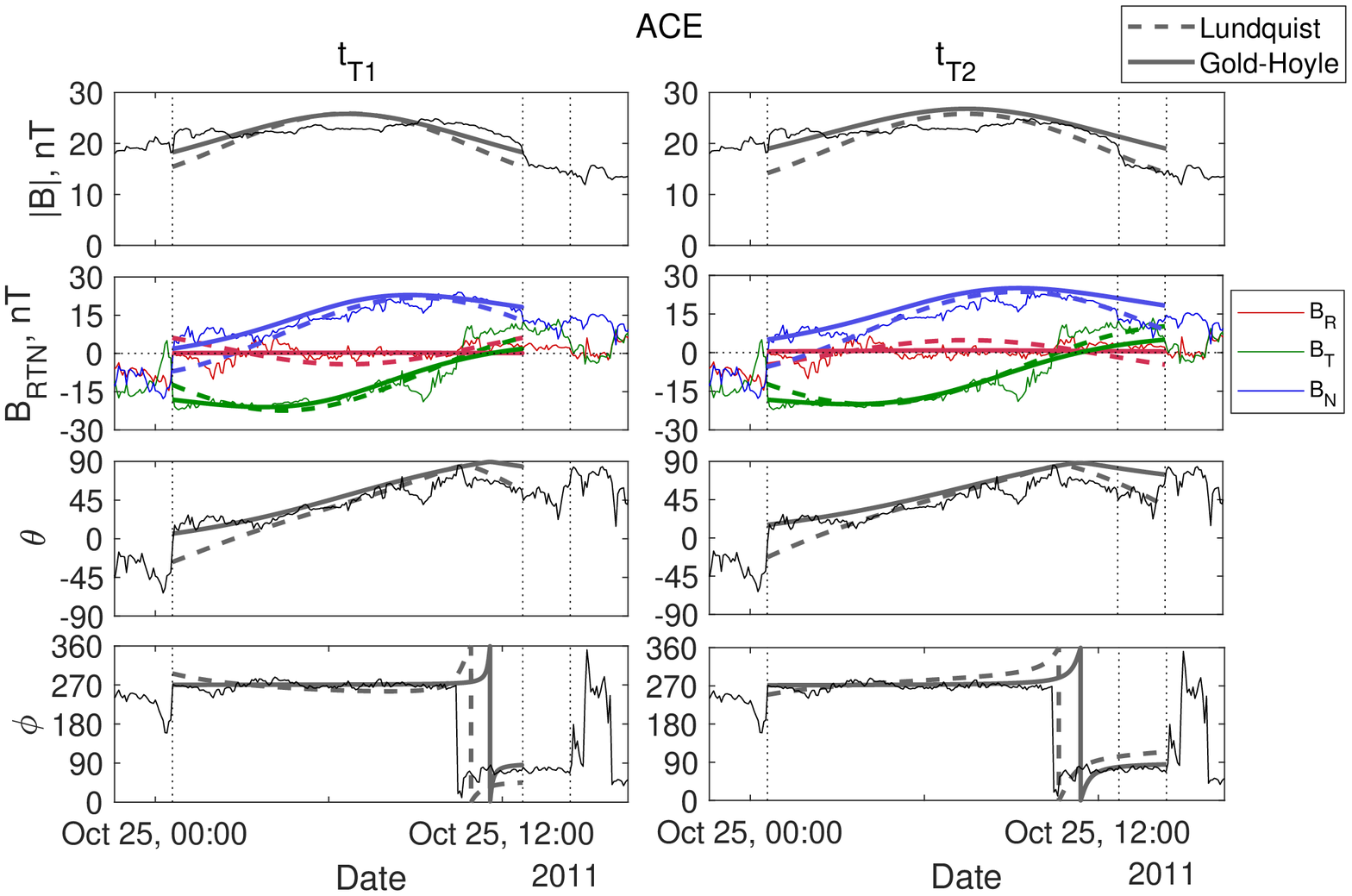}
\caption{Force-free flux rope models fitted to the in situ magnetic field signatures observed by ACE, displayed in the same format as Figure \ref{fig:wind_fits}.}
\label{fig:ace_fits}
\end{figure*}

\begin{figure*}
\centering
\captionsetup{font=small, labelfont=bf}
\includegraphics[width=\textwidth]{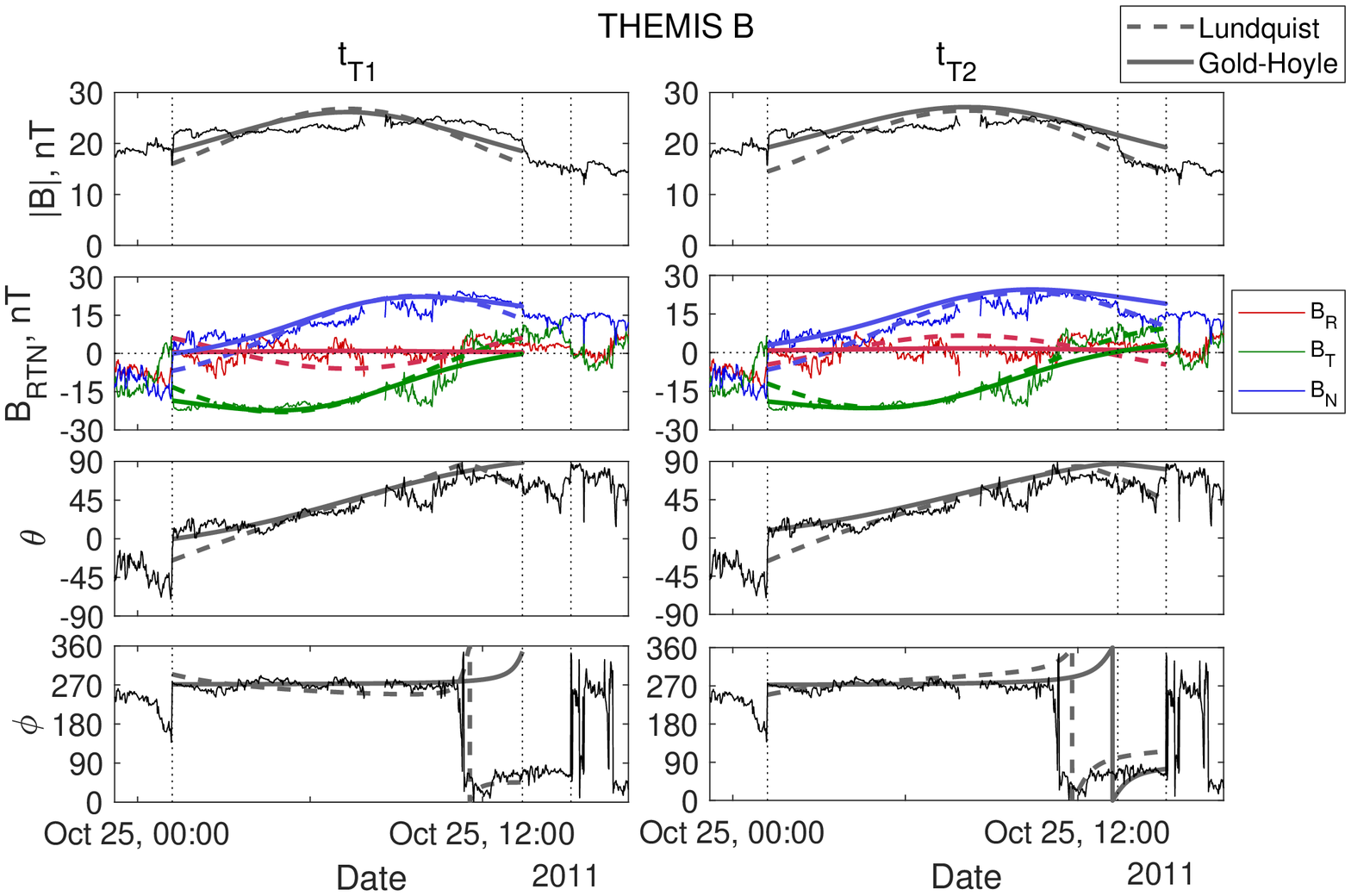}
\caption{Force-free flux rope models fitted to the in situ magnetic field signatures observed by THEMIS B, displayed in the same format as Figure \ref{fig:wind_fits}.}
\label{fig:thb_fits}
\end{figure*}

\begin{figure*}
\centering
\captionsetup{font=small, labelfont=bf}
\includegraphics[width=\textwidth]{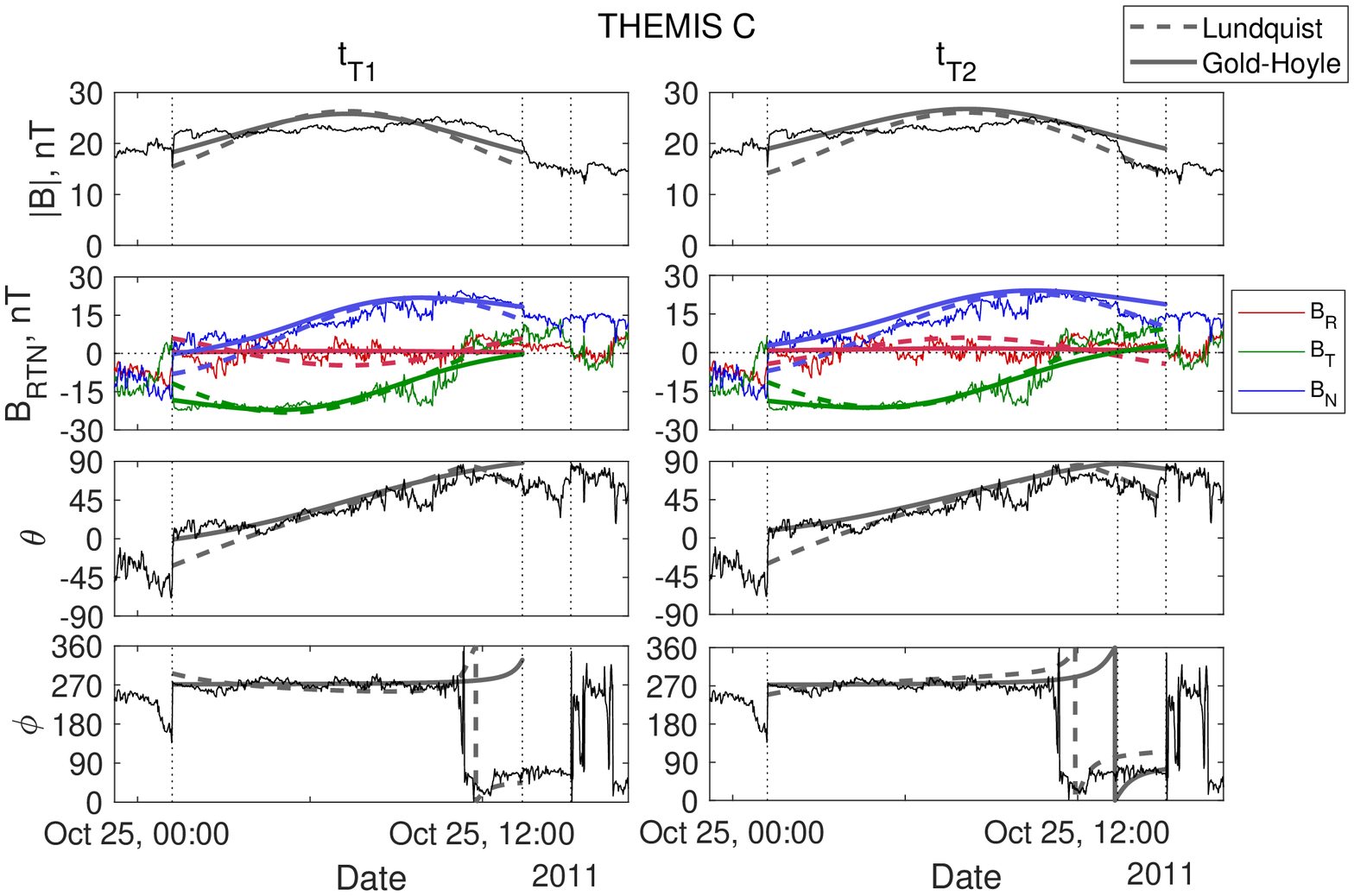}
\caption{Force-free flux rope models fitted to the in situ magnetic field signatures observed by THEMIS C, displayed in the same format as Figure \ref{fig:wind_fits}.}
\label{fig:thc_fits}
\end{figure*}

\end{document}